\documentclass[aps,pre,twocolumn ,superscriptaddress]{revtex4-1}  
\usepackage{graphicx,epsfig}  
\usepackage{dcolumn}   
\usepackage{bm}        
\usepackage{amssymb}   
\usepackage{color}
\usepackage{mathtools}
\usepackage{float}
\usepackage{makecell}
\usepackage[section]{placeins}

\begin{document}

\title{Laser Wakefield Acceleration in a Capillary Gas Cell Producing GeV-Scale High-Quality Electron Beams}

\author{Srimanta Maity}
\email{srimantamaity96@gmail.com}
\affiliation{Atomic, Molecular and Optical Physics Division, Physical Research Laboratory, Navrangpura, Ahmedabad 380009, India}

\author{Francesco Massimo}
\affiliation{Laboratoire de Physique des Gaz et des Plasmas, CNRS, Université Paris-Saclay, 91405 Orsay, France}

\author{Alex Whitehead}
\affiliation{ELI Beamlines Facility, The Extreme Light Infrastructure ERIC, Za Radnicí 835, 25241 Dolní Břežany, Czech Republic}
\affiliation{Czech Technical University in Prague, FNSPE, Břehová 78/7, 11519 Prague, Czech Republic}

\author{Pavel Sasorov}
\affiliation{ELI Beamlines Facility, The Extreme Light Infrastructure ERIC, Za Radnicí 835, 25241 Dolní Břežany, Czech Republic}

\author{Alexander Molodozhentsev}
\affiliation{ELI Beamlines Facility, The Extreme Light Infrastructure ERIC, Za Radnicí 835, 25241 Dolní Břežany, Czech Republic}







\begin{abstract}

\noindent 
 
Laser Wakefield Acceleration (LWFA) is a promising approach for producing high-brightness electron beams in the GeV energy range, offering significant potential for compact next-generation accelerator facilities. In this work, we present a computational study of LWFA in a specially designed single-stage capillary gas-cell target aimed at producing high-quality, GeV-class electron beams. The capillary cell includes a short ($\sim 2$ mm) injection region at the entrance filled with a helium (He) and nitrogen (N$_2$) gas mixture. This is followed by a longer ($\sim 14$ mm) pure He section, which provides the required acceleration length and limits continuous ionization injection, thereby significantly reducing the energy spread of the accelerated beam. Hydrodynamic simulations are performed to optimize the capillary geometry and generate the required two-section gas-pressure profile. The resulting gas-density distributions for various cases are then directly incorporated in Particle-In-Cell (PIC) simulations to study LWFA. In particular, our hydrodynamic simulations demonstrate how tailored density profiles with longitudinal density tapering in the acceleration section can be realized in a capillary gas cell, while the corresponding PIC simulations reveal how these profiles influence the acceleration process and the resulting beam quality. Using a 100~TW-class laser system with parameters relevant to the L2-DUHA laser at the ELI Beamlines Facility, the PIC results demonstrate electron acceleration to mean energies exceeding $1.0$~GeV with high-quality beam properties. Self-injected He electrons are also observed, and their impact on the main beam quality is evaluated. The findings of this study provide valuable insights for upcoming LWFA experiments planned within the EuPRAXIA Project at the ELI Beamlines Facility.

\end{abstract}

\maketitle        
\section{Introduction}\label{intro}

Laser Wakefield Acceleration (LWFA) \cite{Tajima1979, RevModPhys.81.1229} has emerged as one of the most promising approaches for generating high-quality, high-energy electron beams. When an intense laser pulse propagates through a plasma or gas target, its ponderomotive force expels electrons from the laser propagation axis relative to the stationary ions, driving a trailing charge-density wave with intense electromagnetic fields, known as a wakefield. This wakefield propagates with a phase velocity approximately equal to the group velocity of the laser pulse in the medium and exhibits an accelerating gradient on the order of hundreds of Gigavolts per meter (GV/m) \cite{PhysRevLett.70.37, everett1994trapped, modena1995electron}. This accelerating gradient is nearly three orders of magnitude higher than that achievable in conventional radio-frequency accelerators, which are limited by the electrical breakdown threshold of the metallic accelerating cavities. Consequently, electrons injected into this ultra-high-gradient laser-driven wakefield under suitable initial conditions can be rapidly accelerated and transversely focused to produce high-energy electron beams over very short distances. Thus, laser–plasma–based accelerators offer the potential for significantly more compact and cost-effective accelerator technologies \cite{malka2005laser, joshi2007development, hooker2013developments}.  

Following the initial theoretical proposal by Tajima and Dawson in 1979 \cite{Tajima1979}, LWFA has made significant progress over the past few decades toward the experimental realization of high-energy electron acceleration in laser–plasma systems. Early breakthroughs demonstrated the generation of high-energy electron beams from laser–plasma setups \cite{mangles2004monoenergetic, faure2004laser, geddes2004high, leemans2006gev}. These experimental studies were supported by extensive theoretical investigations of the LWFA process \cite{gorbunov1987excitation, sprangle1988laser, bulanov1989excitation, berezhiani1990relativistic, esarey1993optically, bulanov1995two, bulanov1997transverse}. Electron beams with peak energy exceeding 8 GeV have since been achieved using centimeter-scale capillary discharge waveguide \cite{PhysRevLett.122.084801}. More recently, experiments have produced electron beams approaching $\sim 10$ GeV, marking the highest energies reported to date in a single-stage LWFA configuration \cite{aniculaesei2024acceleration, picksley2024matched, li2025longitudinal}. A recent experimental study has also reported the efficient acceleration of plasma electrons driven by a long-wave-infrared laser \cite{zgadzaj2024plasma}. In parallel, numerous theoretical \cite{PhysRevLett.126.104801, PhysRevLett.129.094801, hue2023control, habib2023attosecond, maity2024parametric, guan2025achieving} and experimental studies \cite{PhysRevLett.110.185006, wang2021free, PhysRevLett.126.174801, foerster2022stable, winkler2025active} have focused on improving the quality of the accelerated electron beams to meet the demands of various potential applications \cite{walker2017horizon, emma2021free, vishnyakov2023compact, molodozhentsev:fls2023-th2c2, Whitehead:2024uco, Whitehead:2024dkn, zhou2025compact}.   

The quality of the accelerated electron beams in LWFA depends on several factors, including laser parameters and the parameters of the plasma or gas target. The initial phase of electron injection into the wakefield also affects the quality of the final accelerated beam \cite{maity2025coupling}. Various injection mechanisms, such as self-injection \cite{bulanov1992nonlinear, xu2005electron, PhysRevLett.103.135004, PhysRevLett.103.215006, PhysRevSTAB.15.011302}, density downramp injection \cite{PhysRevE.58.R5257, PhysRevLett.100.215004, ke2021near}, colliding-pulse injection method \cite{PhysRevLett.76.2073, PhysRevLett.79.2682, faure2006controlled, wang2022injection, bohlen2023colliding}, and ionization injection \cite{PhysRevLett.104.025003, PhysRevLett.104.025004, PhysRevLett.105.105003}, have been proposed and demonstrated both theoretically and experimentally.

The ionization-based injection mechanism is considered one of the most effective methods to inject electrons into the wakefield for acceleration. This mechanism takes advantage of the large difference in ionization potential between the inner-shell (K-shell) and outer-shell electrons of high-Z gases (for example, nitrogen). The outer electrons of high-Z gas atoms, and all electrons of low-Z gas atoms, are ionized even at the front part of the laser pulse. In contrast, the K-shell electrons of high-Z gas atoms are ionized only near the central, high-intensity region of the laser pulse. Once these inner-shell electrons are ionized, some of them are trapped in the already formed wake wave created by the previously ionized electrons. In the ionization injection mechanism, the injection mainly occurs near the laser propagation axis. Thus, by properly tuning the laser and plasma parameters, electron beams with low divergence and low emittance can be produced using the ionization-injection mechanism \cite{PhysRevLett.126.174801, steyn2025observation}, whereas these quantities are typically much higher in self-injection and density-downramp injection mechanisms. It also avoids the complexity involved in controlling and synchronizing two driver laser pulses, as required in all-optical injection schemes. 

The main disadvantage of the ionization injection mechanism is the large energy spread of the generated electron beams, which arises from the continuous injection of electrons along the laser propagation distance. Several ideas have been proposed to overcome this issue. For example, the use of an unmatched laser driver has been suggested to self-truncate the ionization-induced injection process \cite{irman2018improved}. Another approach involves using a gas cell setup that separates the injection and acceleration sections \cite{PhysRevLett.126.174801, maity2024parametric, steyn2025observation}. The injection section consists of a mixture of a low-Z gas (hydrogen) and a high-Z gas (nitrogen), while the acceleration section contains pure hydrogen. This method has been both theoretically studied and experimentally verified, producing electron beams with mean energies of $\sim 350$–$400$ MeV and relative energy spreads of $2$–$5\%$ \cite{PhysRevLett.126.104801, PhysRevLett.126.174801, steyn2025observation}. Our present study builds upon this approach to demonstrate the generation of high-quality, GeV-class electron beams using a capillary gas-cell setup.

In this work, we investigate Laser Wakefield Acceleration (LWFA) for generating high-quality, GeV-class electron beams using a truncated ionization-injection mechanism in a single-stage capillary gas cell setup. An inhomogeneous mixture of helium (He) and nitrogen (N$_2$) is considered inside the capillary. Combined hydrodynamic and Particle-In-Cell (PIC) simulations are used to study the conceptual design and the underlying physical processes. To effectively truncate the otherwise continuous ionization-based injection of inner-shell (K-shell) electrons from N$_2$, the capillary is divided into two sections. The first section, which is shorter in length ($\sim 2.0$ mm) and located at the entrance of the capillary, contains a mixture of He and N$_2$ and serves as the injection region. The second section, which is longer ($\sim 12.0$ mm), extends to the capillary exit and contains pure ($100\%$) He, providing the extended acceleration length. Hydrodynamic simulations have been conducted to optimize the design of the capillary setup. Additionally, these simulations provide the gas-pressure profiles along the capillary axis for various configurations. To investigate LWFA using this capillary gas-cell target, the corresponding gas-density profiles for each case are directly incorporated into the PIC simulations. In a recent study \cite{li2025longitudinal}, longitudinal density tapering in gas-jet targets was demonstrated via particle-in-cell simulations and validated experimentally. Their results show that a linearly up-ramped longitudinal density profile can significantly increase the electron beam energy, from 9~GeV to more than 12~GeV. In the present work, our hydrodynamic simulations demonstrate how tailored density profiles, featuring longitudinal upramp or downramp tapering in the acceleration section, can be achieved in a capillary gas cell through precise control of gas-loading parameters and capillary geometry. The corresponding PIC simulations demonstrate how these tapered profiles modify the acceleration dynamics and ultimately influence the final beam quality. A 100~TW-class laser system is considered to drive the wakefield in these gas targets. The PIC simulations accurately capture the injection processes as well as the dynamics of the driver laser pulse and the accelerated electron beams. Our PIC results demonstrate the generation of electron beams with mean energies exceeding $1.0$~GeV, while maintaining high quality upon exiting the capillary.

The remainder of this paper is organized as follows. Section~\ref{hydro_simu_details} describes the hydrodynamic simulations performed in this study. The corresponding results are presented in Section~\ref{hydro_rd}, which discusses the technical aspects of designing the three-channel capillary setup and the gas dynamics inside the capillary. Section~\ref{pic_simu} and its subsections provide the details of the PIC simulations and present the PIC results, highlighting the generation of GeV-class electron beams using LWFA. Finally, Section~\ref{summary} presents a detailed summary of our work along with concluding remarks and future directions.

\section{Hydrodynamic Simulation of Capillary Gas Cell}\label{hydro_simu}

Hydrodynamic simulations were performed using the OpenFOAM \cite{greenshields2025} Computational Fluid Dynamics (CFD) toolbox to obtain the gas density profiles within the capillary gas cell. The main goal was to obtain a gas density profile such that along the capillary channel, there would be a narrow region containing a mixture of nitrogen and helium gases, while the rest of the capillary would be filled only with helium gas, as illustrated in Fig. \ref{fig_sch}(a). This narrow region of mixed gases serves as the injection section for the ionization-injection-based LWFA, whereas the larger helium-filled region acts as an extended acceleration section.

\subsection{Hydrodynamic Simulation Details}\label{hydro_simu_details}

\begin{figure}[hbt!]
  \centering
  \includegraphics[width=3.3in]{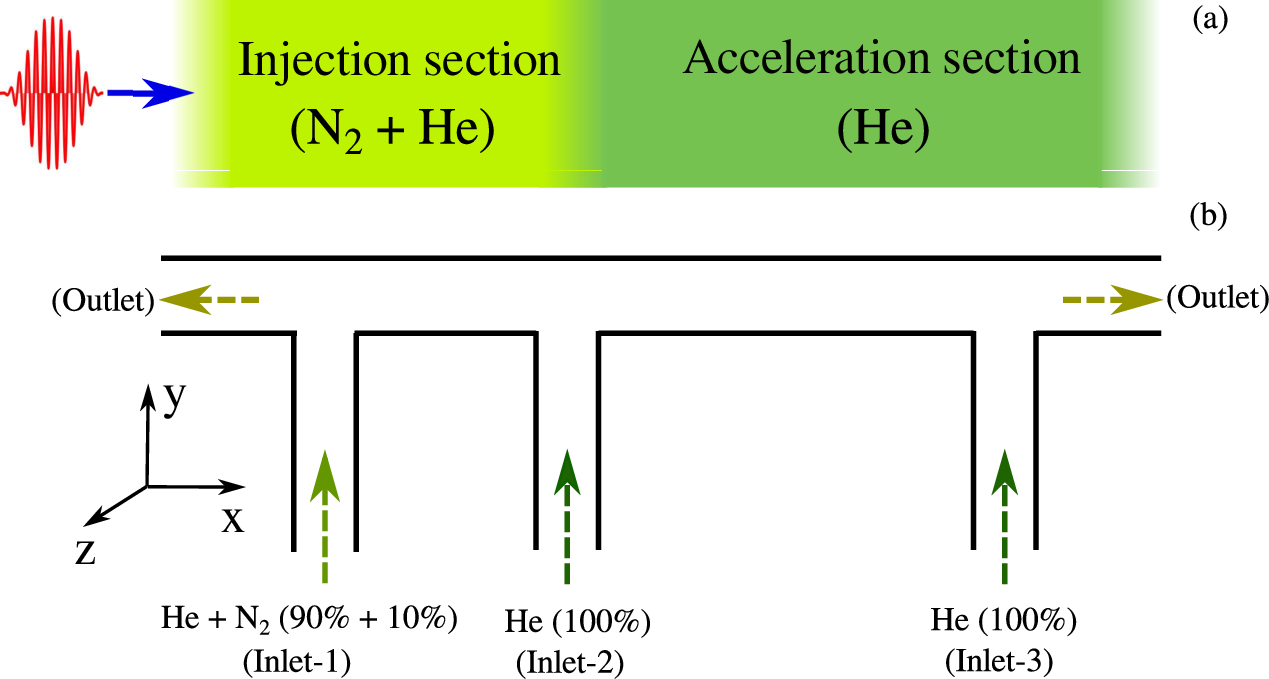}
  \caption{(a)-(b) Schematic of a specially designed capillary gas cell target consisting of two sections: (i) a short injection section with a nitrogen-helium mixture for electron injection, and (ii) a longer acceleration section where the injected electrons are further accelerated. This design helps truncate injection and reduce the final energy spread of the accelerated electron beam.}
\label{fig_sch}
\end{figure}

The capillary setup considered in our study has three inlets (Inlet-1, Inlet-2, and Inlet-3) and two outlets along its axis, as illustrated in Fig. \ref{fig_sch}(b). The axial length (i.e., along $\hat x$) of the square-shaped capillary was considered to be 16 mm, with transverse size in each direction (i.e., $\hat y$ and $\hat z$) of 500 $\mu$m. Additionally, two cubic boxes (pressure ports) with larger volumes (2 mm in each direction) were placed at the capillary entrance (left edge) and exit (right edge) to mimic the experimental vacuum configuration. Square-shaped capillaries are commonly used in LWFA experiments, primarily because they are easier to manufacture and assemble than the circular ones \cite{kruchinin2025experimental}. Inlet-1 is located 1.5 mm from the capillary entrance. Inlet-2 and Inlet-3 are separated by a fixed distance of 10.0 mm, while the separation between Inlet-1 and Inlet-2 was varied by adjusting the position of Inlet-2. Inlets are also considered to be square-shaped gas supply channels with a transverse size of 500 $\mu$m and a length of 5 mm. Inlet-1 was filled with a mixture of nitrogen (N$_2$) and helium (He) gases, with $10\%$ and $90\%$ molar (volume) concentrations, respectively. Inlet-2 and Inlet-3 were filled with pure helium ($100\%$) gas only. The gas flow inside the capillary was driven by the pressure difference between the inlets and outlets. The outlet pressure (i.e., at the vacuum-facing domain walls) was maintained at $1.0 \times 10^{-6}$ mbar. In the simulations, Inlet-1 and Inlet-2 were set to a fixed pressure of 30 mbar, while Inlet-3 was typically set to 40 mbar. In some cases, all inlets were set to 30 mbar.

The gas inside the capillary was modeled as a compressible fluid exhibiting viscous, subsonic flow under turbulent conditions. The flow field was computed by numerically solving the discretized Reynolds-Averaged Navier-Stokes (RANS) equations within the CFD framework. A density-based solver, \textit{multicomponentFluid}, was employed, utilizing first-order Euler time discretization, first-order upwind schemes for convective terms, and second-order Gauss linear schemes for gradients, to provide stable transient simulations of compressible, turbulent, multicomponent flows. A constant heat capacity was assumed, and viscosity was modeled using the Sutherland law with standard parameters for nitrogen and helium obtained from the literature. The turbulent properties of the flow were described using the $k$–$\epsilon$ turbulence model, where $k$ is the turbulent kinetic energy and $\epsilon$ represents the turbulent dissipation rate. However, our hydrodynamic simulation results were found to be nearly identical when using a laminar flow model. This is because the characteristic Reynolds number ($\mathit{Re}$) remains well below the threshold value required (typically $\sim 2000$) for the turbulence flow. For example, the maximum steady-state flow velocity ($U$) of helium gas through a capillary channel with a square cross-section of side length $L = 500~\mu\text{m}$ is observed to be approximately $U \sim 800$~m/s. Using the kinematic viscosity of helium at temperature $T = 300$~K and pressure $P = 40$~mbar, $\nu \sim 3.1 \times 10^{-3}$~m$^{2}$/s, the resulting Reynolds number is $\mathit{Re} = UL/\nu \sim 129$, which is far below the turbulent-flow regime. In our simulations, mutual diffusion between helium and nitrogen was modeled using the \textit{FickianEddyDiffusivity} formulation available in OpenFOAM. The binary diffusivity was set to $D = 1.77\times10^{-3}$ m$^2$/s, calculated using the Chapman-Enskog relation \cite{chapman1990mathematical} for a dilute He-N$_2$ mixture at the operating conditions of $T = 300$ K and $P = 40$ mbar. The turbulent contribution to scalar transport was represented using the standard RANS-based eddy-diffusivity approach. However, under our operating conditions, the influence of molecular diffusion is minimal. For the characteristic length scale $L = 500~\mu\mathrm{m}$ and flow velocity $U \approx 800~\mathrm{m\,s^{-1}}$, the corresponding Péclet number is $\mathrm{Pe} = UL/D \approx 2.2\times10^{2} \gg 1$, indicating that convective transport strongly dominates over diffusive mixing in this regime. The initial simulation time step was set to $10$ ns. An "adjustTimeStep" scheme was employed, with a maximum Courant number of 0.9, allowing the time step to vary dynamically according to the local velocity field.
 
The capillary cell used in this study is described using the Cartesian coordinate system. The capillary axis is aligned with the $x$-direction, while the vertical and horizontal transverse directions correspond to the $y$- and $z$-axes, respectively, as shown in Fig.~\ref{fig_sch}(b). Variations along the $z$-axis, arising from the front and back walls of the capillary, are assumed to be negligible compared to those in the $x$–$y$ plane. Based on this assumption, an \textit{empty} boundary condition was applied along the $z$-direction, which eliminates any variation, flux, or flow in that direction and allows the simulation to focus only on the $x$–$y$ cross-section, reducing computational cost. The simulation domain was discretized using a \textit{hexahedral} mesh generated with the OpenFOAM \textit{blockMesh} utility, with finer resolution near the inlet and outlet regions. The average mesh size in the $x$-$y$ plane was approximately $0.01 \times 0.01$~mm$^2$. We have also performed an additional hydrodynamic simulation for a particular case (Case-1) by increasing the number of grid points in the transverse plane to test numerical convergence. The comparative results shown in the Supplemental Material (Fig. S1) indicate that the pressure profile remains unchanged as the grid resolution is increased.

The initial boundary condition for the pressure field was specified as \textit{fixedValue} at the inlets, while at the outlets it was defined as \textit{waveTransmissive} with a target pressure of $1.0 \times 10^{-6}$~mbar. This configuration effectively minimizes spurious wave reflections from the outlet patches. At the walls, a \textit{zeroGradient} boundary condition was applied to the pressure field. For the velocity field, the \textit{pressureInletVelocity} boundary condition was employed at the inlets, whereas the \textit{pressureInletOutletVelocity} condition was applied at the outlets, ensuring a pressure-driven flow of gases through the inlets and outlets. At the walls, a \textit{noSlip} boundary condition was applied to the velocity field to model the viscous interaction between the gas and the solid surfaces. A constant room temperature of 300~K was considered in the simulation.


\subsection{Hydrodynamic Simulation Results and Discussion}\label{hydro_rd}

\begin{figure*}
  \centering
  \includegraphics[width=0.9\textwidth]{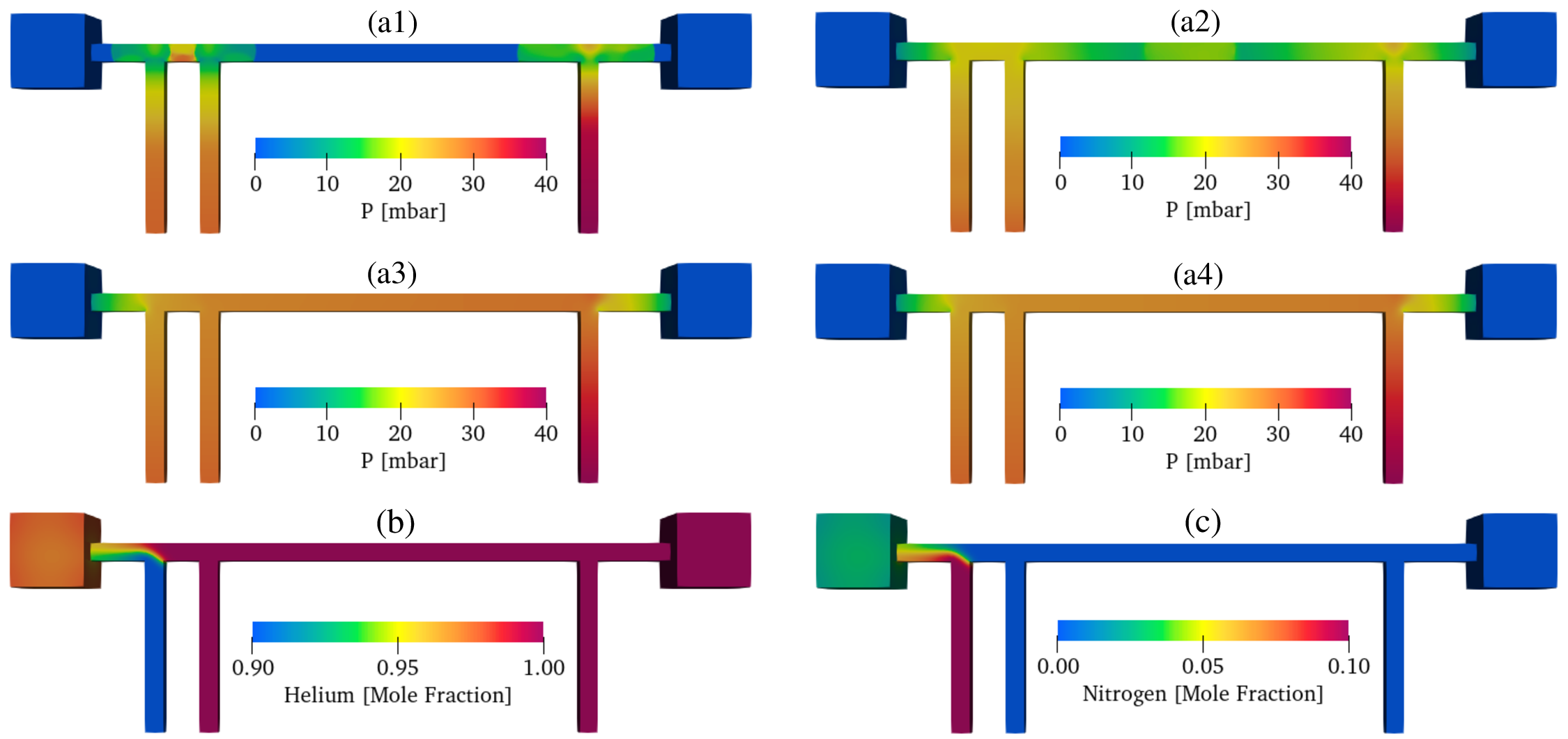}
  \caption{(a1)–(a4) Snapshots of the gas-filling process in the capillary cell for a particular case (Case-2) at simulation times $t = 100~\mu\text{s}$, $150~\mu\text{s}$, $250~\mu\text{s}$, and $400~\mu\text{s}$, respectively. Subplots (b) and (c) show the mole fractions of helium and nitrogen, respectively, at $t = 400~\mu\text{s}$. Here, Inlet-1 is supplied with a $90\%$ He and $10\%$ N$_2$ mixture, while Inlet-2 and Inlet-3 are fed with $100\%$ He.}
\label{fig_2d_openfoam}
\end{figure*}

\begin{figure*}
  \centering
  \includegraphics[width=0.9\textwidth]{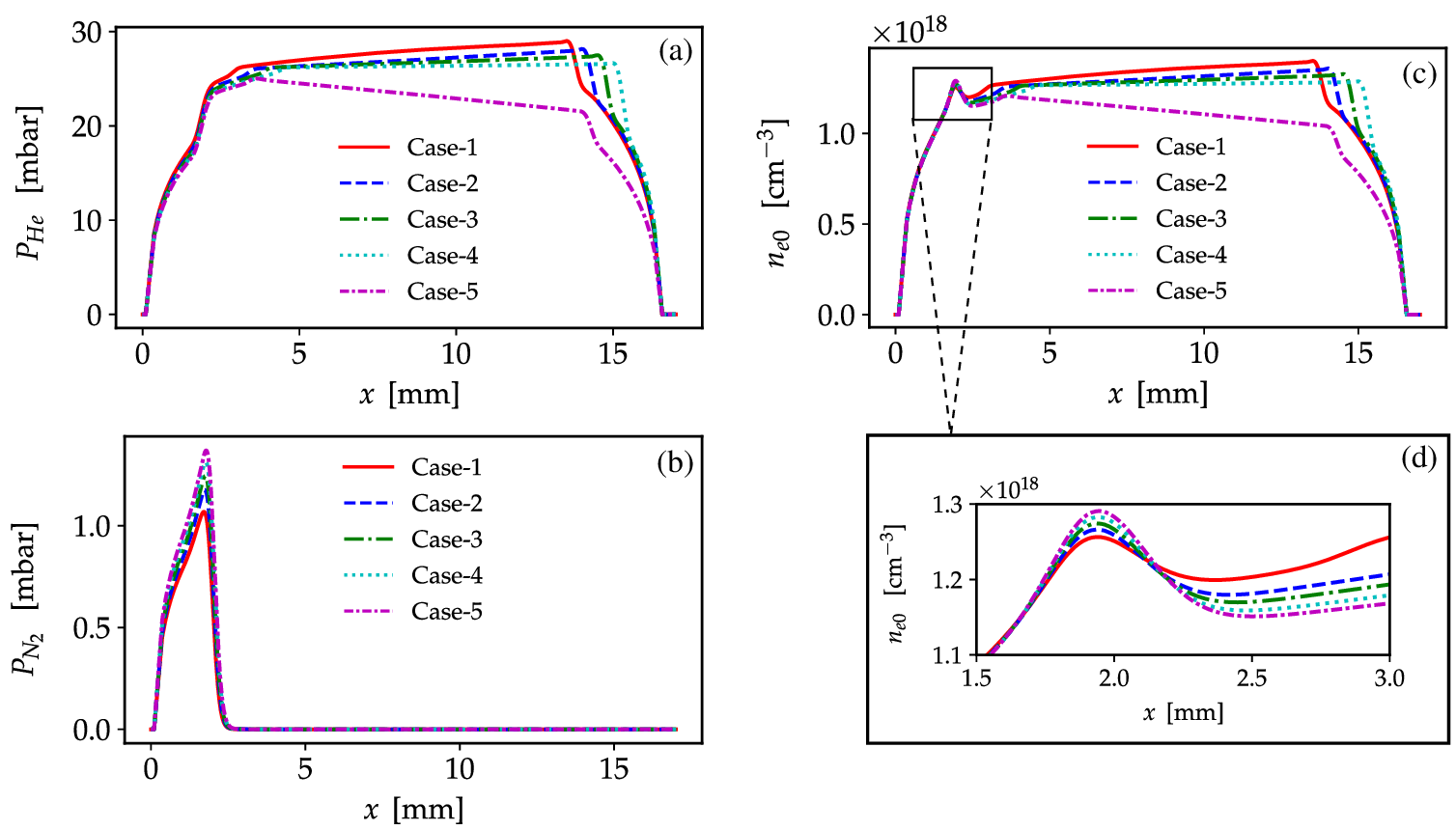}
  \caption{(a)–(b) On-axis partial pressure profiles of He and $N_2$ along the capillary axis. The different cases are as follows: For Case-1 to Case-4, the separation between Inlet-1 and Inlet-2 is 0.5 mm, 1.0 mm, 1.5 mm, and 2.0 mm, respectively. In these cases, Inlet-1 and Inlet-2 are maintained at 30 mbar, while Inlet-3 is set to 40 mbar. For Case-5, all three inlets are maintained at 30 mbar, but the separation between Inlet-1 and Inlet-2 is 1.0 mm. The pressure at both outlets is fixed at $1.0 \times 10^{-6}$ mbar (vacuum) for all cases. (c)–(d) Corresponding on-axis electron density profiles calculated from the partial pressures, assuming two electrons per He atom and ten electrons per $N_2$ molecule. These on-axis 1D profiles are extracted at the steady-state condition at a simulation time of $t = 400~\mu\mathrm{s}$.
}
\label{fig_density_openfoam}
\end{figure*}

In this section, we present the hydrodynamic simulation results for different cases. In Fig. \ref{fig_2d_openfoam}(a1)-(a4), snapshots of the gas filling process inside the capillary are shown at simulation times $t = 100~\mu\mathrm{s}$, $150~\mu\mathrm{s}$, $250~\mu\mathrm{s}$, and $400~\mu\mathrm{s}$, respectively. Specifically, the pressure field within the capillary is displayed at four different times during the simulation for a selected case. In this case, the distance between Inlet-1 and Inlet-2 was 1.0~mm. Inlet-1 was supplied with a mixture of $90\%$ He and $10\%$ N$_2$, driven by a fixed inlet pressure of 30 mbar. Inlet-2 and Inlet-3 were supplied with $100\%$ He, with fixed inlet pressures of 30 mbar and 40 mbar, respectively. The gas-filling process was completed after approximately 200~$\mu\mathrm{s}$, at which point the pressure inside the capillary became nearly constant, indicating that a steady state had been reached. The simulations were then continued up to $t = 400~\mu\mathrm{s}$ to ensure stable steady-state conditions. It is observed that the pressure inside the capillary cell is slightly lower than the inlet pressures (i.e., 30 and 40 mbar), consistent with the findings of a previous study reported in Ref. \cite{bagdasarov2022discharge}. This pressure reduction is attributed to the finite viscosity of the gas and the no-slip boundary condition at the surfaces, which prevents the free flow of gas through the supply channel and capillary. 

In Fig. \ref{fig_2d_openfoam}(b)–(c), the mole fraction distributions of He and N$_2$ at the final steady state ($t = 400~\mu\mathrm{s}$) are shown. It is observed that near Inlet-1, particularly toward the entrance of the capillary, helium exhibits a mole fraction of approximately 0.9, while nitrogen has a mole fraction of about 0.1. Nitrogen is confined to this narrow region near the inlet, whereas the remaining portion of the capillary channel is predominantly filled with pure helium gas. The confinement of nitrogen near Inlet-1 is attributed to the fact that the gas entering through this inlet experiences a stronger pressure-gradient force toward the left outlet (the capillary entrance) than toward the right outlet (the capillary exit), due to the shorter distance to the left. At later times, the higher or more uniformly distributed helium pressure generates a convective flux that opposes the diffusion of nitrogen towards the central region of the capillary channel. As a result, the transport of nitrogen is limited to a narrow area near its inlet (i.e., Inlet-1), while the remainder of the capillary is dominated by helium.

Hydrodynamic simulations were performed for five separate cases. In Case-1 to Case-4, the separation between Inlet-1 and Inlet-2 was varied to 0.5 mm, 1.0 mm, 1.5 mm, and 2.0 mm, respectively. In all four of these cases, Inlet-1 and Inlet-2 were maintained at a pressure of 30 mbar,  while Inlet-3 was set to 40 mbar. For Case-5, the separation between Inlet-1 and Inlet-2 was fixed at 1.0 mm, and in this case, all three inlets were set to a consistent pressure of 30 mbar. The separation between Inlet-2 and Inlet-3 was held constant at 10.0 mm for all five cases.

The capillary geometry and inlet configuration were chosen to obtain a plasma density profile suitable for GeV-level laser wakefield acceleration. In particular, the gas pressure and the separation between the inlets were selected such that the resulting average plasma density lies in the range $1.2\times10^{18}$--$1.4\times10^{18}$ cm$^{-3}$, corresponding to a gas pressure of approximately 30--40 mbar as obtained from our hydrodynamic simulations. This density range was estimated using LWFA scaling laws \cite{lu2007generating} so that the dephasing ($L_d$) and pump depletion ($L_{pd}$) lengths are comparable to the effective acceleration length of the capillary. For the chosen laser parameters, the corresponding values are approximately $L_d \sim 12$ mm and $L_{pd} \sim 12.5$ mm at $n_e \sim 1.4 \times 10^{18}$ cm$^{-3}$. In the simulations, the separation between Inlet-2 and Inlet-3 is fixed at 10 mm, which approximately defines the acceleration length and ensures that the main beam does not reach the dephasing limit during the acceleration process.

The axial pressure profiles within the capillary for the five cases are presented in Fig.~\ref{fig_density_openfoam}. These data are extracted from the simulations at steady-state ($t = 400~\mu\mathrm{s}$). It should be noted that, in the hydrodynamic simulations, in addition to the 16 mm long capillary channel, an additional 2 mm vacuum region is present on each side of the capillary. However, in Fig.~\ref{fig_density_openfoam}, we show a shorter vacuum region on each side of the capillary: 0.1 mm before the capillary entrance and 0.5 mm after the capillary exit. This choice was made to reduce the computational cost of the PIC simulations, where these profiles are directly used, as described later. It is observed that the partial pressure of helium (P$_{He}$) increases approximately linearly along the capillary axis for Case-1 to Case-4, as shown in Fig. \ref{fig_density_openfoam}(a). This behavior can be attributed to the higher pressure at Inlet-3 (40 mbar) relative to Inlet-1 and Inlet-2 (30 mbar). In all cases, it is observed that the maximum pressure within the capillary channel remains below 30 mbar, which is lower than the inlet pressures (30 and 40 mbar). This result is attributed to the effect of finite gas viscosity, as discussed earlier. It is also noteworthy that in Case-5, where all three inlets are maintained at a fixed pressure of 30 mbar, a linearly decreasing (down-ramp) helium pressure profile is observed along the capillary channel. This behavior is expected because Inlet-1 and Inlet-2 are much closer to each other (1.0 mm separation) than Inlet-2 and Inlet-3 (which are separated by 10 mm). Consequently, a higher net gas feed occurs near the left side of the capillary (entrance) compared to the right side (exit).

The most prominent effect of increasing the separation between Inlet-1 and Inlet-2 (from Case-1 to Case-4) is observed in the amount of nitrogen fed into the capillary channel for a fixed nitrogen concentration ($10\%$) through Inlet-1. The axial nitrogen pressure profiles along the capillary channel for all five cases are presented in Fig. \ref{fig_density_openfoam}(b). In all cases, the nitrogen gas is confined within a narrow region extending approximately 2 mm from the capillary entrance. The peak pressure occurs at the location where Inlet-1 connects to the capillary channel. It is further observed that, as the separation distance between Inlet-1 and Inlet-2 increases (from Case-1 to Case-4), the peak value of the nitrogen partial pressure also increases, indicating a greater amount of nitrogen being fed into the capillary channel. This behavior can be attributed to the reduction in the helium convective pressure flow from Inlet-2 with increasing separation distance, which allows more nitrogen to enter the capillary. For a similar reason, in Case-5, when the helium pressure at Inlet-3 is reduced to 30 mbar, a larger amount of nitrogen is able to accumulate inside the capillary channel. 

These outcomes are expected to have a significant impact on laser-driven wakefield acceleration (LWFA) in such a capillary configuration. In particular, the injected electron beam charge (and consequently, the beam current) as well as other beam quality parameters are likely to be affected. The results of Particle-In-Cell (PIC) simulations of LWFA incorporating these pressure profiles for various cases are presented and discussed in the following sections.


\section{Particle-In-Cell (PIC) Simulation of Laser Wakefield Acceleration}\label{pic_simu}


\subsection{PIC Simulation Details}\label{pic_simu_details}

We performed Particle-In-Cell (PIC) simulations to investigate laser wakefield acceleration (LWFA) of electrons in the capillary setup, as described in Section~\ref{hydro_simu}. The simulations were carried out using the open-source PIC code SMILEI \cite{derouillat2018smilei}. The gas target parameters used in the PIC simulations were directly adopted from the hydrodynamic simulation results presented in Section~\ref{hydro_rd}.

To reduce computational cost, we assumed that helium is fully ionized and nitrogen atoms are ionized up to the fifth ionization state, i.e., only the two K-shell electrons remain bound. This assumption is reasonable because helium’s two electrons and the five outer-shell electrons of nitrogen have relatively low ionization potentials and are readily ionized by the leading edge of the laser pulse. In contrast, the two tightly bound K-shell electrons of nitrogen require much higher intensities and are ionized only near the peak of the laser field. The resulting electron density profiles, calculated under these assumptions, are shown in Fig.~\ref{fig_density_openfoam}(c) for the five cases discussed in Section~\ref{hydro_rd}. In all cases, a distinct density hump is observed, originating from the contribution of five electrons per nitrogen atom. A zoomed view of the net axial electron density near the density transition region is presented in Fig.~\ref{fig_density_openfoam}(d). It is evident that, from Case-1 to Case-5, the amplitude of the density hump increases, consistent with the trends observed in Fig.~\ref{fig_density_openfoam}(b). 

Accordingly, in our PIC simulations, we initially considered three species of particles: (i) electrons ionized from helium atoms, with a longitudinal density profile derived from Fig.~\ref{fig_density_openfoam}(a) using the ideal gas law at room temperature (300 K); (ii) the five outer-shell electrons from each nitrogen atom; and (iii) nitrogen ions in the N$^{5+}$ charge state. The longitudinal density profiles for species (ii) and (iii) were obtained from Fig.~\ref{fig_density_openfoam}(b).

We have also performed an additional simulation considering initially neutral species for both gases and compared the results with the pre-ionized case. The results are presented in the Supplemental Material (Fig. S2). The comparison indicates that there is no significant change in the laser propagation dynamics between these two cases.

The PIC simulations were performed using the quasi-3D axisymmetric cylindrical geometry implemented in SMILEI, with a decomposition into two azimuthal modes. A moving window scheme propagating at a speed of $c$ (the speed of light in vacuum) was employed. The simulation window had a longitudinal extent (along $\hat{x}$) of $82\lambda$ and a transverse (radial, $\hat{y}$) radius of $140\lambda$. Here, $\lambda = 0.82~\mu\text{m}$ is the wavelength of the laser used in this study. The grid resolution was set to $\lambda/50$ in the longitudinal direction and $\lambda/5$ in the radial direction, with an integration timestep of $\lambda/(51c)$. 

All three species mentioned earlier were treated as macroparticles, initially cold and randomly initialized within each cell. Eight particles per cell were considered for each species. The longitudinal ($\hat{x}$) density profiles used in the simulations are shown in Fig.~\ref{fig_density_openfoam}(b)–(c). In the radial direction ($\hat{y}$), no variation in density was considered. 

A terawatt-class laser (100~TW on target) was launched from the left boundary of the simulation box, propagating along $\hat{x}$ with linear polarization along $\hat{y}$. The laser was assumed to have Gaussian profiles in both the longitudinal and transverse directions. The laser parameters used in the simulation are as follows: wavelength of $820~\text{nm}$, pulse energy of $3.5~\text{J}$, pulse duration (FWHM) of $35~\text{fs}$, and beam waist (radius) of $20~\mu\text{m}$. With these parameters, the laser intensity at vacuum focus is $I_0 \approx 1.6 \times 10^{19}~\text{W/cm}^2$, corresponding to a normalized vector potential of $a_0 = 2.8$. The laser was initially focused at a position $x = 4.5~\text{mm}$ from the left boundary of the simulation box. The tunnel ionization model based on the Ammosov-Delone-Krainov direct current (ADKdc) ionization rate \cite{ammosov1986tunnel}, along with zero-momentum initialization for newly ionized electrons, was employed to account for the laser-induced ionization of the inner-shell (K-shell) electrons of N$^{5+}$ ions.

The laser parameters in the PIC simulations were fixed to correspond to the L2-DUHA laser system at the ELI Beamlines Facility. Other simulation parameters were chosen such that a beam charge in the range of 30-50 pC can be obtained, which is suitable for potential applications such as incoherent or coherent photon radiation in the extreme ultraviolet or X-ray regime. To reduce the dimensionality of the parameter space, the nitrogen concentration was fixed at $10\%$ with respect to the helium pressure at Inlet-1. Several preliminary simulations were conducted for different laser focusing positions. Based on these results, the focus position was set to $x \sim 4.5$ mm to achieve the desired charge range. The PIC simulation results for a particular case (Case-1), considering three different laser focusing positions, are presented in the Supplemental Material (Fig. S3).


    




\subsection{PIC Simulation Results and Discussion}\label{pic_rd}

\begin{figure*}
  \centering
  \includegraphics[width=0.9\textwidth]{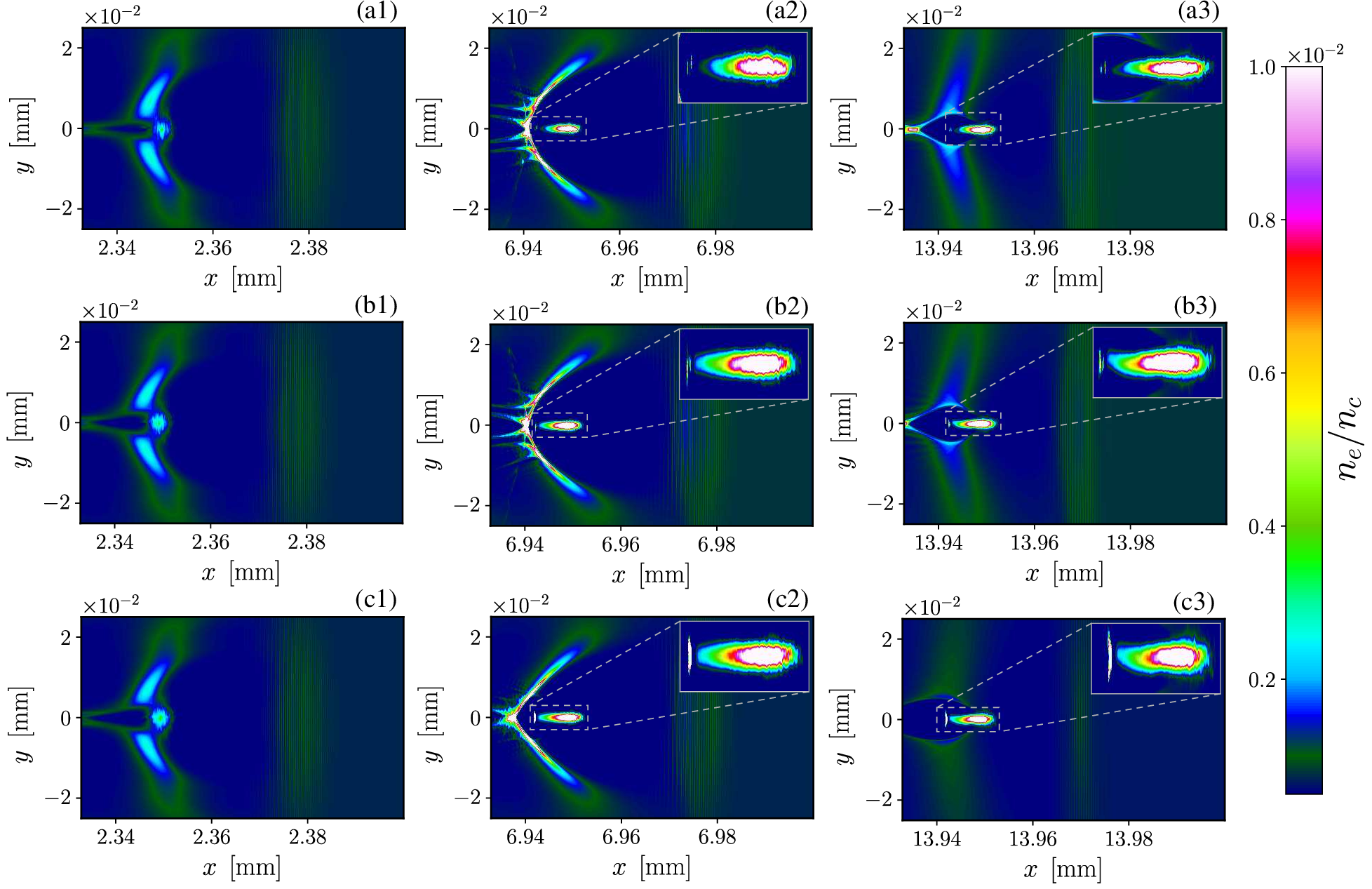}
  \caption{Electron density distributions on the 2D x–y plane at three different simulation times for (a1–a3) Case-2, (b1–b3) Case-4, and (c1–c3) Case-5, respectively. The electron density is normalized to the critical density $n_c$ corresponding to the laser frequency used in the simulation.}
\label{fig_pic_dens2d}
\end{figure*}

\begin{figure*}
  \centering
  \includegraphics[width=0.9\textwidth]{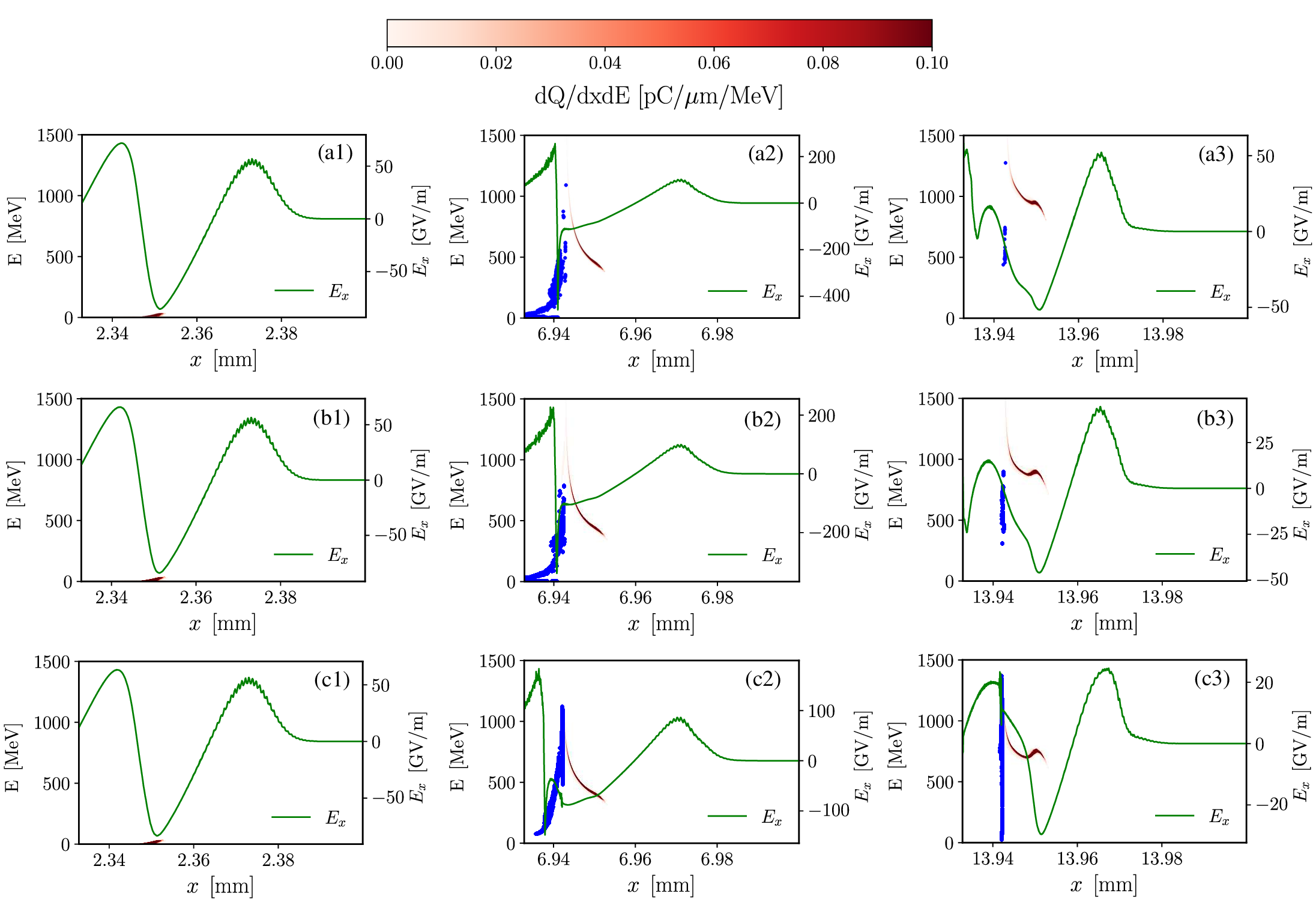}
  \caption{On-axis longitudinal electric field $E_x$ and the corresponding energy distributions of accelerated electrons at three different simulation times are shown for (a1)–(a3) Case-2, (b1)–(b3) Case-4, and (c1)–(c3) Case-5, respectively. The longitudinal electric field $E_x$ is plotted as green solid lines. The color-mapped patches (see colorbar) represent electrons ionized from N$^{5+}$ ions, while blue scattered dots correspond to electrons originating from helium with energies exceeding 5.0 MeV.}
\label{fig_pic_phase_space}
\end{figure*}

\begin{figure}
  \centering
  \includegraphics[width=3.3in]{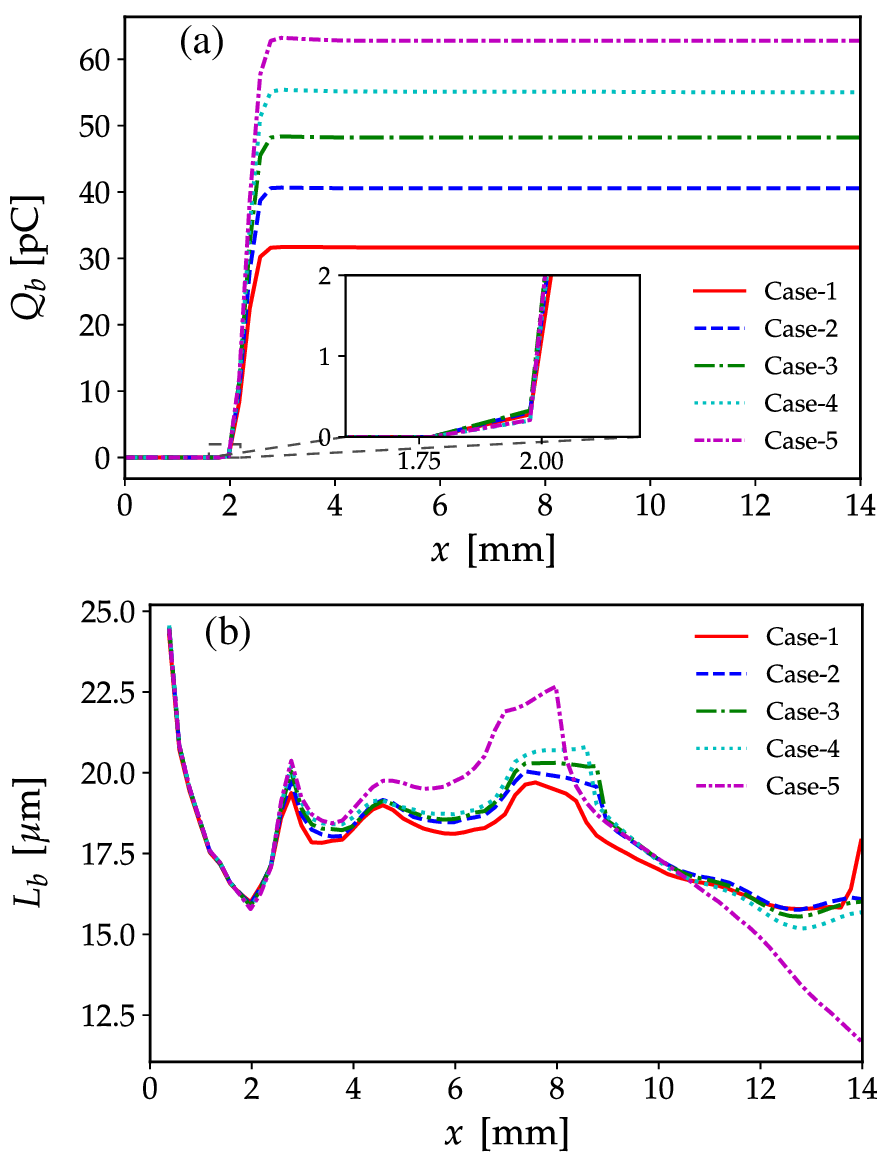}
  \caption{(a) Evolution of the net charge ($Q_b$) of electrons ionized from the K-shell of N$^{5+}$ ions as a function of laser propagation distance. Only electrons with energies greater than 5.0 MeV are considered. (b) Evolution of bubble length ($L_b$) of the first wake behind the laser pulse as a function of propagation distance.}
\label{fig_pic_Q_lb_x}
\end{figure}

\begin{figure}
  \centering
  \includegraphics[width=3.3in]{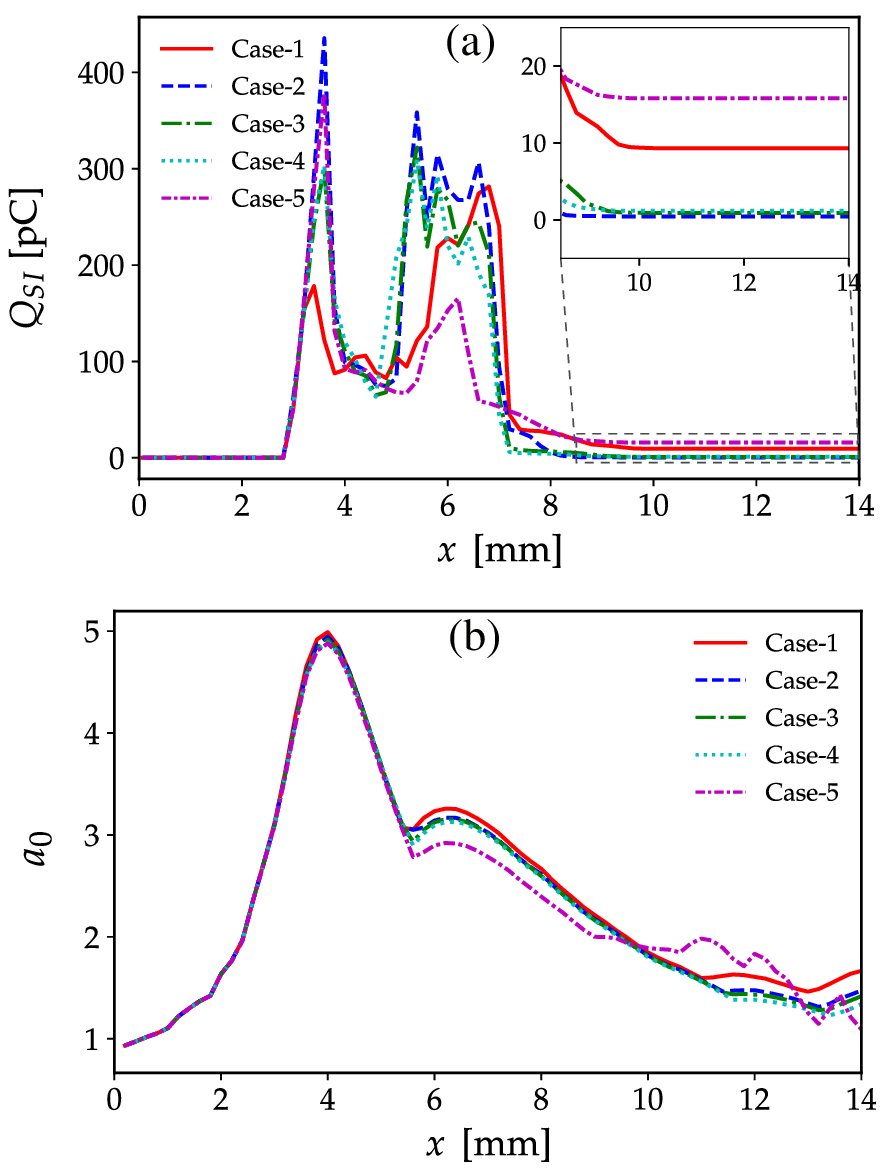}
  \caption{(a) Evolution of the net charge of self-injected electrons from He as a function of laser propagation distance. Only electrons with energies greater than 5.0 MeV are considered. (b) Variation of the peak normalized vector potential ($a_0$) of the laser pulse with propagation distance.
}
\label{fig_pic_Q_a0}
\end{figure}

\begin{figure}
  \centering
  \includegraphics[width=3.3in]{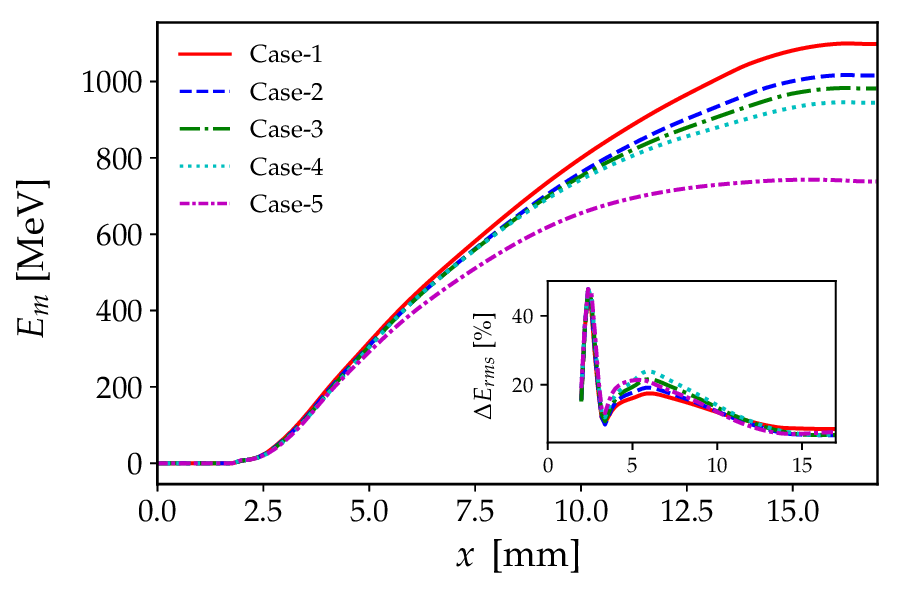}
  \caption{Mean energy ($E_m$) of the ionization-injected electron beam as a function of laser propagation distance. The inset shows the corresponding RMS energy spread.}
\label{fig_pic_Em_x}
\end{figure}

\begin{figure*}
  \centering
  \includegraphics[width=0.9\textwidth]{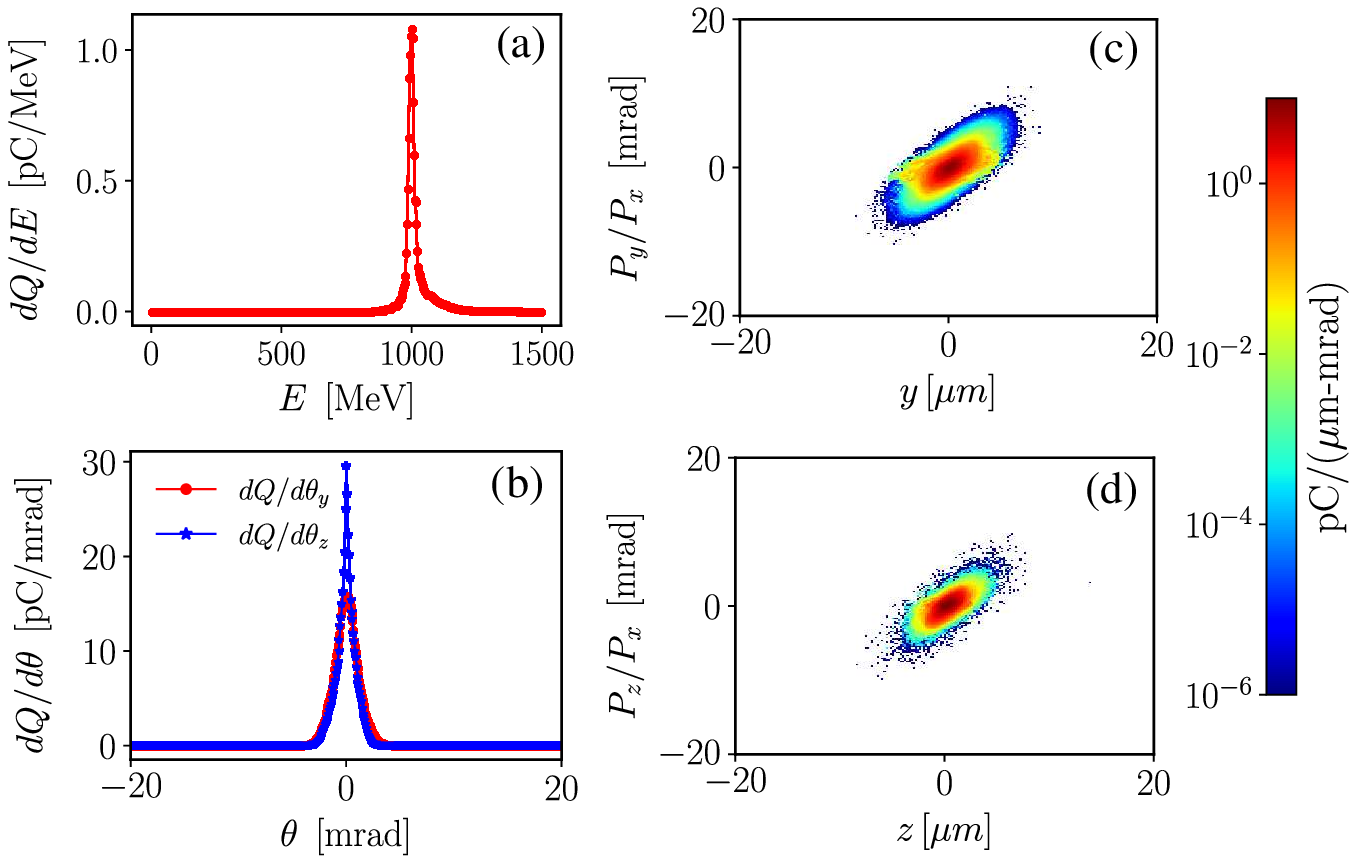}
  \caption{Properties of the accelerated electron beam at the exit of the gas cell, i.e., $x = 16.5$ mm, for a particular case (Case-2) with a separation of 1.0 mm between Inlet-1 and Inlet-2. In this case, Inlet-1 and Inlet-2 are fixed at 30 mbar, while Inlet-3 is maintained at 40 mbar.}
\label{fig_pic_beam_diag_final}
\end{figure*}

Snapshots of the electron density distribution in 2D x-y plane at three different simulation times are shown in Fig. \ref{fig_pic_dens2d}: (a1)–(a3) for Case-2, (b1)–(b3) for Case-4, and (c1)–(c3) for Case-5. From Fig. \ref{fig_pic_dens2d}(a1)–(c1), it can be seen that in all cases, electrons are injected and trapped in the first wake wave behind the laser pulse, starting at approximately $x \approx 2.2$ mm. The evolution of the injected electrons and the wakefield (i.e., the first wake wave) is illustrated in Fig. \ref{fig_pic_dens2d}(a2)–(c3), showing that the injected electron beams remain confined within the first wake wave behind the laser pulse. The overall size of the injected beam is largest for Case-5, followed by Case-4 and Case-2. A more detailed view of Fig. \ref{fig_pic_dens2d}(a2)–(c2) highlights the presence of a second electron beam located just behind the primary beam, also trapped within the wakefield. This feature becomes more pronounced in Fig. \ref{fig_pic_dens2d}(a3)–(c3). Furthermore, the presence of the secondary beam is most prominent in Case-5, followed by Case-4, and is least evident in Case-2. The two beams are distinctly separated, indicating that they are injected at different phases of the wakefield through two separate mechanisms.

To characterize the origin of the injected electrons, we track electrons from all three species considered in our study: electrons corresponding to the first five ionization states of nitrogen atoms, electrons originating from helium atoms, and electrons that are ionized from the K-shell of N$^{5+}$ ions. The phase space distribution of these electrons, along with the longitudinal wake electric field $E_x$, is illustrated in Fig. \ref{fig_pic_phase_space} for three different simulation times across three cases: (a1)-(a3) for Case-2, (b1)-(b3) for Case-4, and (c1)-(c3) for Case-5. 

The color patches shown in Fig. \ref{fig_pic_phase_space} illustrate the longitudinal phase-space distribution of the ionized K-shell electrons (refer to the colorbar). In this analysis, we consider only the electrons with energies greater than 5.0 MeV. It is observed that the ionization-injected electrons are trapped within the accelerating and focusing phase of the electric field of the first wake wave, allowing them to continue gaining energy as the laser propagates. After a propagation distance of $ x\simeq 14 $ mm, these electrons achieve a mean energy of approximately 1.0 GeV for Case-2, about 0.95 GeV for Case-4, and around 0.75 GeV for Case-5, as depicted in Fig. \ref{fig_pic_phase_space}(a3)-(c3). We have also observed that electrons from helium are self-injected into the wakefield. This is illustrated in Fig. \ref{fig_pic_phase_space}, where the blue scattered dots represent the self-injected helium electrons with energies exceeding 5.0 MeV. We have observed that the self-injection occurs near the laser focusing point, much later compared to the ionization injection, as can be seen from Fig. \ref{fig_pic_phase_space}(a1)-(c1), where no helium electrons are injected yet. 

Our simulations indicate that electrons corresponding to the first five ionization states of nitrogen are not injected into the wakefield. Instead, they only contribute to the formation of the wake wave. This conclusion is based on the fact that we do not observe any such electrons with energies above $5.0$ MeV. These electrons are located in the region $x \sim 0.01$ mm to $x \sim 2.2$ mm (see Fig. \ref{fig_density_openfoam}(b)), which is well before the laser focal point at $x = 4.5$ mm. As a result, the laser intensity in this region is not sufficient to drive either self-injection or density down-ramp injection. However, the central part of the laser pulse still has sufficient intensity to ionize the inner K-shell electrons of the N$^{5+}$ ions. Once these K-shell electrons are ionized near the center of the laser pulse, they become situated within the fully formed wake wave. Some of these ionized electrons, which meet certain criteria \cite{PhysRevLett.104.025003}, become trapped in the wake wave and gain energy. To explicitly identify the injection mechanism of the inner-shell (K-shell) electrons of nitrogen, we track the net charge of electrons ionized from N$^{5+}$ ions as a function of propagation distance, as shown in Fig. \ref{fig_pic_Q_lb_x}(a). Our analysis indicates that injection begins at $x \sim 1.75$ mm, which occurs before the density downramp region (between $ x \sim 2.0$ mm and $2.4$ mm), as shown in the inset of Fig. \ref{fig_pic_Q_lb_x}(a). It should be noted that only electrons with energies exceeding $5.0$ MeV are considered in this analysis. Therefore, the actual injection location is expected to occur earlier than $x\sim 1.75$ mm. This confirms that the initial electron injection is primarily driven by the ionization-induced mechanism. However, a sudden increase in beam charge is observed between $x \sim 2.0$ mm and $2.2$ mm. This enhancement is caused by the density transition in this region (see Fig. \ref{fig_density_openfoam}(d)), which reduces the wake phase velocity due to a rapid increase in the bubble length, as shown in Fig. \ref{fig_pic_Q_lb_x}(b). Therefore, the injection mechanism in our study can be described as downramp-assisted ionization injection \cite{thaury2015shock}. Furthermore, the net charge of the ionization-injected electrons remains constant throughout the propagation, indicating no beam charge loss.

Our analysis reveals that the self-injected electrons from helium are trapped in the same plasma wake as the ionization-injected electrons. However, they remain well separated in phase space, as shown in Fig. \ref{fig_pic_phase_space}(a2)–(c3). The self-injected electrons produce a strong beam-loading effect, which significantly modifies the accelerating field $E_x$, as illustrated in Fig. \ref{fig_pic_phase_space}(a2)–(c2). However, since the self-injected electrons are located behind the main ionization-injected beam throughout the propagation, the wakefield perturbation caused by their beam loading is confined to the region trailing the main beam. Moreover, both electron beams propagate at approximately the wake phase velocity ($\sim c$), and, due to causality, perturbations in the wakefield generated by trailing charges cannot influence the fields experienced by leading charges. Consequently, the beam-loading effect of the self-injected electrons does not affect the acceleration process or beam quality of the ionization-injected electron beam. Our simulations also indicate that most of these self-injected electrons decelerate and defocus over time, eventually being lost later in the propagation, as shown in Fig. \ref{fig_pic_phase_space}(a3)–(c3). This occurs because the bubble length steadily decreases in each case after $x \sim 8$ mm, as shown in Fig. \ref{fig_pic_Q_lb_x}(b).

To quantitatively describe the evolution of self-injected helium electrons, we calculate the net charge ($Q_{SI}$) of these self-injected electrons as a function of laser propagation distance. In our analysis, we consider only electrons with energies greater than 5.0 MeV. The results are presented in Fig. \ref{fig_pic_Q_a0}(a) for all five cases. It is observed that self-injection begins in all cases after $x\sim 3.0$ mm. In each case, there are two distinct locations where the net charge of the self-injected electrons reaches peak values. The peak values of the self-injected electrons vary from $Q_{SI} = 150$ to 400 pC across the five cases. These two peaks in the net charge profile correspond to two peaks observed in the profile of the normalized vector potential $a_0$, as shown in Fig. \ref{fig_pic_Q_a0}(b). 

The laser power ($P_0 = 100$ TW) considered in our study exceeds the critical power ($P_c = 17\omega_0^2/\omega_{pe}^2 = 22.5$ TW). Here, $\omega_0$ is the laser frequency and $\omega_{pe}$ is the plasma frequency, calculated based on an average plasma density of $n_e = 1.25\times 10^{24}$ m$^{-3}$. As a result, the driver laser beam undergoes the relativistic self-focusing effect as it propagates through the plasma. Consequently, $a_0$ reaches its peak value at $ x \sim 3.8$ mm, prior to the laser focusing point, achieving a maximum value of 5.0, which is significantly higher than the vacuum focus value of ($a_0 = 2.8$). Additionally, the laser pulse considered in our study is initially not matched with the plasma density, i.e., $k_pw_0 \neq 2\sqrt{a_0}$ \cite{lu2007generating}, where $k_p$ and $w_0$ represent the wave number of a plasma wave and the laser spot size. As a result, the diverging effect of the laser pulse does not counterbalance the self-focusing effect, leading to oscillation of the laser spot and the emergence of a second peak in the $a_0$ profile, which can be seen in Fig. \ref{fig_pic_Q_a0}(b). 

It is interesting to notice that for Case-1, the net charge of the self-injected electrons is significantly lower at the first peak compared to the second peak, despite a higher value of $a_0$ for the first peak. This feature distinctly differs from the trend observed in other cases, as illustrated in Fig. \ref{fig_pic_Q_a0}(a). This difference can be attributed to the fact that, in Case-1, the rate of increase in the density around $x \sim 3.0$ mm is significantly higher than in the other cases (see Fig. \ref{fig_density_openfoam}(d)). This leads to a rapid decrease in wake size with propagation distance, resulting in less injection of helium electrons near the first peak. Moreover, in each case, we see a steady drop in the self-injected trapped electron charge after the second peak. This again matches the evolution of $a_0$, which decreases quickly after $x \sim 8$ mm. As $a_0$ decreases during propagation, the wake size also becomes smaller, as illustrated in Fig. \ref{fig_pic_Q_lb_x}(b). Because of this, most of the self-injected electrons dephase and move into the decelerating and defocusing phase of the second wake wave. 

After a propagation distance of $x \sim 9$ mm, the net charge of the self-injected electrons becomes nearly constant for all cases, with values ranging from 0.1 pC to 18 pC, as shown in the inset of Fig. \ref{fig_pic_Q_a0}(a). The highest remaining charge corresponds to Case-5, owing to the largest amount of self-injected charge in this case. For Case-2 to Case-4, the final saturated charge lies in the range of 0.1 pC to 0.5 pC, with the lowest value observed in Case-2, followed by Case-3, and the highest in Case-4. This trend can be attributed to the rate of density increase, which is fastest in Case-2, slower in Case-3, and slowest in Case-4. Consequently, a larger fraction of self-injected helium electrons in Case-2 undergo dephasing and enter the decelerating and defocusing phase of the second wake wave. However, Case-1 does not follow this trend, despite having an upramp density profile. In this case, the final net charge of the self-injected electrons is $Q_{SI} \sim 12$ pC, which is significantly higher than in Case-2 to Case-4. This can be attributed to the comparatively higher plasma density and $a_0$ at later stages of the evolution, which leads to more rapid acceleration of the self-injected electrons. As a result, fewer self-injected electrons undergo phase slippage into the second wake wave. 

Overall, our simulation results show that, in addition to the primary ionization-injected electron beam originating from the K-shell of nitrogen, a large number of self-injected electrons from helium are also accelerated. The simulations further indicate that these two electron beams remain well separated within the wake structure throughout the propagation. Due to causality, the trailing self-injected beam cannot influence the dynamics of the leading ionization-injected beam. In most cases, the energy spectra of the self-injected beams are clearly separated from those of the primary beam (see Fig.~\ref{fig_pic_phase_space}(a3)-(b3)). Therefore, in principle, these self-injected electrons can be experimentally filtered out using a dipole magnet. However, in certain cases (e.g., Case-5; see Fig.~\ref{fig_pic_phase_space}(c3)), the energy spectrum of the self-injected electrons overlaps with that of the ionization-injected beam. In LWFA experiments, such spectral overlap makes it difficult to distinguish between the two electron bunches and can significantly degrade the quality of the final accelerated electron beam. Importantly, our simulations demonstrate that using a gas target with a slowly rising density ramp in the accelerating section (as in Case-2, Case-3, and Case-4) can strongly suppress, or even completely eliminate, self-injected electrons during propagation.

The mean energy gained by electrons ionized from the K-shell of N$^{5+}$ ions is shown in Fig.~\ref{fig_pic_Em_x}. It is observed that for Case-1 to Case-4, the ionization-injected electrons gain energy monotonically during propagation until reaching the exit downramp, after which the energy saturates. The rate of energy gain, and consequently the final energy achieved, increases from Case-4 to Case-1. In contrast, for Case-5, the mean energy of the injected electrons almost saturates well before reaching the exit downramp. In all cases, the injected electron beams do not reach the dephasing limit, as shown in Fig. \ref{fig_pic_phase_space}(a3)-(c3). Although $a_0$ steadily decreases after a propagation distance of $\sim 8$ mm in each case, the tailored increasing density profile for Case-1 to Case-4 prevents the wakefield amplitude from decreasing at the same rate. As a result, the injected electrons continue to gain energy as they propagate up to the exit downramp, with a higher rate of energy gain for cases with a stronger increasing density tailoring. For Case-5, however, both $a_0$ and the plasma density decrease with the axial position. Consequently, the electrons stop gaining energy well before reaching the exit ramp and the dephasing distance. The corresponding RMS energy spread is shown in the inset. The energy spread follows the same trend across all cases. The variation in the energy spread as a function of propagation distance appears to follow the evolution of $a_0$, which strongly influences the beam's position within the phase of the wakefield.

The properties of the accelerated electron beam, initially injected through ionization injection from the inner-shell (K-shell) electrons of N$^{5+}$ ions, are shown in Fig. \ref{fig_pic_beam_diag_final} at the exit of the capillary gas cell (i.e., at $x=16.5$ mm). The results correspond to Case-2, where Inlet-1 and Inlet-2 were separated by 1.0 mm and both were maintained at an inlet pressure of 30 mbar, while Inlet-3 was located 10 mm downstream of Inlet-2 and was held at 40 mbar. Figure \ref{fig_pic_beam_diag_final}(a) presents the energy distribution of the generated electron beam, showing a mean energy of approximately $1.0$ GeV with a narrow energy spread. The angular distributions (beam divergence) in the transverse $y-$ and $z-$directions are shown in Fig. \ref{fig_pic_beam_diag_final}(b). The distributions are centered at zero and have narrow widths of about 1.5~mrad, which confirms that electrons move primarily along the beam axis with only small transverse momentum. The nearly identical shapes in $y$ and $z$ indicate that the focusing fields experienced by the beam electrons are close to uniform in both transverse planes. The transverse phase-space distributions shown in Fig. \ref{fig_pic_beam_diag_final}(c)-(d) exhibit tilted ellipses in the $P_{y}$-$y$ and $P_{z}$-$z$ planes. This indicates a weak correlation between the transverse position and momentum, implying that the beam is not fully emittance matched. Such phase-space coupling may originate from asymmetric radial focusing fields in the plasma wake and/or from the off-axis injection of ionized electrons, both of which introduce a position-dependent transverse momentum. However, the transverse momentum spreads are still negligible compared to the longitudinal momentum, so the beam remains well collimated.

\begin{table*}
\begin{tabular}{|c|c|c|c|c|c|}
\hline
    \makecell{Beam Properties} & \makecell{Case-1} & \makecell{Case-2} & \makecell{Case-3} & \makecell{Case-4} & \makecell{Case-5} \\ 
\hline
\makecell{Beam charge} & 31 pC & 41 pC & 48 pC & 55 pC & 63 pC \\ 
\hline
\makecell{Mean energy} & 1.1 GeV & 1.02 GeV & 0.98 GeV & 0.94 GeV & 0.74 GeV \\ 
\hline
\makecell{Relative energy\\ spread (FWHM)} & 2.2 $\%$ & 2.4 $\%$ & 3.1 $\%$ & 4.6 $\%$ & 2.2 $\%$ \\ 
\hline
\makecell{Beam charge\\ within FWHM} & 10 pC & 21 pC & 30 pC & 33 pC & 17 pC \\
\hline
\makecell{RMS beam divergence\\ ($\theta_{y,z}$)} & (1.1, 0.8)~mrad & (1.1, 0.9)~mrad & (1.2, 1.0)~mrad & (1.3, 1.0)~mrad & (1.5, 1.2)~mrad \\ 
\hline
\makecell{RMS normalized\\ emittance ($\epsilon_{ny,nz}$)} & (2.1, 1.1)~mm.mrad & (2.1, 1.2)~mm.mrad & (2.1, 1.2)~mm.mrad & (2.2, 1.3)~mm.mrad & (2.2, 1.2)~mm.mrad \\ 
\hline
\makecell{RMS beam size\\ ($\sigma_{x,y,z}$)} & (1.6, 1.1, 0.8)~$\mu$m & (1.7, 1.0, 0.7)~$\mu$m & (1.9, 1.1, 0.8)~$\mu$m & (2.0, 1.1, 0.8)~$\mu$m  & (2.1, 1.2, 0.9)~$\mu$m \\ 
\hline
\end{tabular}
\caption{Properties of the ionization-induced injected electron beam at the exit of the gas cell target (i.e., at $x = 16.5$ mm) for different cases. The laser was focused at $x = 4.5$ mm in all cases.}
\label{table_beam_params}
\end{table*}

We have analyzed the properties of the electron beam produced by ionization injection for all cases at the exit of the capillary gas cell ($x = 16.5$~mm). These properties are listed in detail in Table~\ref{table_beam_params} for each case. In our study, the generated electron beam has a charge ranging from 30 to 65~pC. From Case-1 to Case-5, the net beam charge increases. This follows from Fig. \ref{fig_density_openfoam}(b), which shows that the amount of nitrogen seeded into the capillary cell increases from Case-1 to Case-5. Additionally, the density transition becomes steeper (see Fig. \ref{fig_density_openfoam}(d)) as we move from Case-1 to Case-5. The final energy gain of the accelerated electron beam ranges from $1.1$~GeV to $0.74$~GeV, from Case-1 to Case-5, respectively. The decrease in the final energy from Case-1 to Case-5 is mainly due to the increase in beam charge from Case-1 to Case-5. A higher beam charge flattens the local gradient of the accelerating wake electric field because of the space-charge electric field of the beam. This effect was also reported in Ref. \cite{maity2024parametric}. In addition, the axial plasma density profile in the accelerating section of the capillary is not uniform. For Case-1 to Case-4, the density has a linear up-ramp, and at any fixed axial position the density decreases from Case-1 to Case-4. Case-5 instead has a down-ramp profile, as shown in Fig. \ref{fig_density_openfoam}(c). Since the amplitude of the accelerating wake electric field follows the scaling $E_x \propto \sqrt{n_e}$, a higher plasma density produces a stronger accelerating field. Thus, for a fixed acceleration length, electrons gain more energy in Case-1, followed by Case-2, Case-3, and Case-4, with the least energy gain in Case-5. In our study, the relative FWHM energy spread of the accelerated electron beam varies between $2.2\%$ and $4.6\%$ across the different cases. From Case-1 to Case-4, the relative energy spread increases from $2.2\%$ to $4.6\%$, primarily due to the corresponding increase in beam charge. In Case-5, although the net beam charge is comparatively high (63 pC), the charge contained within the FWHM is lower (17 pC), leading to a reduced relative energy spread. The beam divergence, RMS normalized emittance, and the RMS beam size remain nearly constant across all cases as presented in Table~\ref{table_beam_params}. 

Among the configurations investigated in our study, Case-2 demonstrates the most favorable performance, producing an electron beam with a net charge of 41 pC, a mean energy of 1.01 GeV, a relative FWHM energy spread of $2.4\%$, an RMS beam divergence of 1.0 mrad, and an RMS normalized emittance of 2.1 mm·mrad and 1.2 mm·mrad in the two transverse planes, respectively. A further optimization in designing the capillary geometry, laser parameters, and gas target parameters can reduce the energy spread of the generated electron beam to below $1\%$. For example, periodic changes in the accelerating field slope, as theoretically proposed in a recent study \cite{guan2025achieving}, can be implemented in our capillary setup to reduce the electron-beam energy spread to below the percentage level.

\section{Summary and Conclusions}
\label{summary}

Laser wakefield acceleration (LWFA), generating high-quality electron beams with energy in the GeV range, has been investigated using computational modeling. In particular, the capability of a capillary gas cell target to produce high-quality electron beams at the GeV scale has been extensively examined. A start-to-end modeling combining hydrodynamic and Particle-In-Cell (PIC) simulations has been performed. The ionization injection mechanism has been employed for the LWFA. 

In this study, a single gas cell target with two separate sections is considered. The first section, about $\sim 2$ mm long, contains a gas mixture of $90\%$ helium and $10\%$ nitrogen. The second section, with a longer length of $\sim 14$ mm, consists of pure helium gas. The purpose of such a design is to limit the ionization-induced injection process from the high-Z gas atoms of nitrogen within the first section, effectively reducing the energy spread of the generated electron beams, which is otherwise large in the ionization-based injection mechanism. The second section, which consists of pure helium gas, then provides a long acceleration region for the electrons injected in the first section. 

Extensive hydrodynamic simulations have been performed to accurately design such a capillary gas cell target. We have demonstrated that a capillary cell with three inlets, where one inlet is supplied with a gas mixture of nitrogen and helium, and the other two inlets are fed with pure helium gas, is highly effective in creating a gas cell target that separates two sections. The hydrodynamic simulation results, including the capillary geometry, inlet and outlet pressures, gas density profiles, and other parameters, have been analyzed and discussed in detail. LWFA using this specially designed capillary cell has been investigated by directly incorporating the gas density profiles obtained from hydrodynamic simulations into the PIC simulations. Our hydrodynamic modeling reveals how tailored gas-density profiles, with upramp or downramp density tapering in the acceleration section, can be produced in a single-stage capillary setup, while PIC simulations highlight how these profiles affect the injection and acceleration processes, ultimately determining the quality of the accelerated beams.

A 100 TW class laser system with a wavelength of 820 nm, a pulse duration of 35 fs, a beam waist of 30 $\mu$m, and a pulse energy of 3.5 J has been considered to drive the wakefield. These laser parameters correspond to the L2-DUHA Laser system of the ELI Beamlines Facility. The PIC simulation results show that electrons injected through ionization-based injection can reach a mean energy of 1.1 GeV after propagating 14 mm through the capillary cell. For the optimized case, the generated electron beam has a net charge of $\sim 40$ pC, a relative energy spread of $\sim 2.2\%$, an RMS beam divergence of $1.0$ mrad, and an RMS normalized emittance of $2.1$ mm·mrad and $1.2$ mm·mrad in the two transverse planes, respectively. With further optimization of the capillary design and laser parameters, the energy spread of the generated electron beams can be reduced to below $1\%$. In addition to the ionization-injected electrons, our simulations also capture helium electrons injected via self-injection. These electrons trail the primary ionization-injected beam throughout the propagation, and, due to causality, they cannot affect the dynamics of the primary beam. However, in some cases, their energy spectra can overlap, leading to a degradation of the overall beam quality to be observed in experiments. As a possible solution, our simulations indicate that employing a slowly increasing density profile in the accelerating section can control, or even eliminate, such self-injected electrons. 

The methods presented in this work will soon be applied in upcoming experimental campaigns at the ELI Beamlines Facility under the EuPRAXIA Project \cite{walker2017horizon, 10.1063/5.0286730}. The findings of this study will be highly valuable for those future experiments.



\section*{Acknowledgments}

 S. M. acknowledges support from the INSPIRE Faculty Fellowship (DST/INSPIRE/04/2024/000118; Faculty Registration No. IFA24-PH 329) of the Department of Science $\&$ Technology, Government of India. The computations were performed on the Param Vikram-1000 High Performance Computing Cluster of the Physical Research Laboratory (PRL). A. W.  acknowledges support from the European Union's Horizon Europe research and innovation programme under grant agreement no. 101073480 and the UKRI guarantee funds. This work benefited from the HPC resources of CINES under the allocation A0170510062 (VirtualLaplace) made by GENCI. This work also benefited from the use of the meso-scale HPC “3Lab Computing” hosted at École polytechnique and administrated by the Laboratoire Leprince-Ringuet, Laboratoire des Solides Irradiés and Laboratoire pour l’Utilisation des Lasers Intenses.

 


\section*{References}
\bibliographystyle{unsrt}
\bibliography{ref.bib}

@article{Tajima1979,
  title = {Laser Electron Accelerator},
  author = {Tajima, T. and Dawson, J. M.},
  journal = {Phys. Rev. Lett.},
  volume = {43},
  issue = {4},
  pages = {267--270},
  numpages = {0},
  year = {1979},
  month = {Jul},
  publisher = {American Physical Society},
  doi = {10.1103/PhysRevLett.43.267},
  url = {https://link.aps.org/doi/10.1103/PhysRevLett.43.267}
}

@article{RevModPhys.81.1229,
  title = {Physics of laser-driven plasma-based electron accelerators},
  author = {Esarey, E. and Schroeder, C. B. and Leemans, W. P.},
  journal = {Rev. Mod. Phys.},
  volume = {81},
  issue = {3},
  pages = {1229--1285},
  numpages = {0},
  year = {2009},
  month = {Aug},
  publisher = {American Physical Society},
  doi = {10.1103/RevModPhys.81.1229},
  url = {https://link.aps.org/doi/10.1103/RevModPhys.81.1229}
}

@article{PhysRevLett.70.37,
  title = {Ultrahigh-gradient acceleration of injected electrons by laser-excited relativistic electron plasma waves},
  author = {Clayton, C. E. and Marsh, K. A. and Dyson, A. and Everett, M. and Lal, A. and Leemans, W. P. and Williams, R. and Joshi, C.},
  journal = {Phys. Rev. Lett.},
  volume = {70},
  issue = {1},
  pages = {37--40},
  numpages = {0},
  year = {1993},
  month = {Jan},
  publisher = {American Physical Society},
  doi = {10.1103/PhysRevLett.70.37},
  url = {https://link.aps.org/doi/10.1103/PhysRevLett.70.37}
}

@article{everett1994trapped,
  title={Trapped electron acceleration by a laser-driven relativistic plasma wave},
  author={Everett, M. and Lal, A. and Gordon, D. and Clayton, C. E. and Marsh, K. A. and Joshi, C.},
  journal={Nature},
  volume={368},
  number={6471},
  pages={527--529},
  year={1994},
  publisher={Nature Publishing Group UK London},
  doi = {10.1038/368527a0},
  url = {https://doi.org/10.1038/368527a0}
}

@article{modena1995electron,
  title={Electron acceleration from the breaking of relativistic plasma waves},
  author={Modena, A. and Najmudin, Z. and Dangor, A. E. and Clayton, C. E. and Marsh, K. A. and Joshi, C. and Malka, V. and Darrow, C. B. and Danson, C. and Neely, D. and Walsh, F. N.},
  journal={Nature},
  volume={377},
  number={6550},
  pages={606--608},
  year={1995},
  publisher={Nature Publishing Group UK London},
  doi = {10.1038/377606a0},
  url = {https://doi.org/10.1038/377606a0}
}

@article{malka2005laser,
  title={Laser-plasma accelerators: a new tool for science and for society},
  author={Malka, V. and Faure, J. and Glinec, Y. and Lifschitz, A. F.},
  journal={Plasma physics and controlled fusion},
  volume={47},
  number={12B},
  pages={B481},
  year={2005},
  publisher={IOP Publishing},
  doi = {10.1088/0741-3335/47/12B/S34},
  url = {http://dx.doi.org/10.1088/0741-3335/47/12B/S34}
}

@article{joshi2007development,
  title={The development of laser-and beam-driven plasma accelerators as an experimental field},
  author={Joshi, C.},
  journal={Physics of plasmas},
  volume={14},
  number={5},
  pages={525},
  year={2007},
  publisher={American Institute of Physics},
  doi = {10.1063/1.2721965},
  url = {https://doi.org/10.1063/1.2721965}
}

@article{hooker2013developments,
  title={Developments in laser-driven plasma accelerators},
  author={Hooker, S. M.},
  journal={Nature Photonics},
  volume={7},
  number={10},
  pages={775--782},
  year={2013},
  publisher={Nature Publishing Group UK London},
  doi = {10.1038/nphoton.2013.234},
  url = {https://doi.org/10.1038/nphoton.2013.234}
}

@article{mangles2004monoenergetic,
  title={Monoenergetic beams of relativistic electrons from intense laser--plasma interactions},
  author={Mangles, S. P. D. and Murphy, C. D. and Najmudin, Z. and Thomas, A. G. R. and Collier, J. L. and Dangor, A. E. and Divall, E. J. and Foster, P. S. and Gallacher, J. G. and Hooker, C. J. and others},
  journal={Nature},
  volume={431},
  number={7008},
  pages={535--538},
  year={2004},
  publisher={Nature Publishing Group UK London},
  doi = {10.1038/nature02939},
  url = {https://doi.org/10.1038/nature02939}
}

@article{faure2004laser,
  title={A laser--plasma accelerator producing monoenergetic electron beams},
  author={Faure, J. and Glinec, Y. and Pukhov, A. and Kiselev, S. and Gordienko, S. and Lefebvre, E. and Rousseau, J. -P. and Burgy, F. and Malka, V.},
  journal={Nature},
  volume={431},
  number={7008},
  pages={541--544},
  year={2004},
  publisher={Nature Publishing Group UK London},
  doi = {10.1038/nature02963},
  url = {https://doi.org/10.1038/nature02963}
}

@article{geddes2004high,
  title={High-quality electron beams from a laser wakefield accelerator using plasma-channel guiding},
  author={Geddes, C. G. R. and T{\'o}th, C. and Van Tilborg, J. and Esarey, E. and Schroeder, C. B. and Bruhwiler, D. and Nieter, C. and Cary, J. and Leemans, W. P.},
  journal={Nature},
  volume={431},
  number={7008},
  pages={538--541},
  year={2004},
  publisher={Nature Publishing Group UK London},
  doi = {10.1038/nature02900},
  url = {https://doi.org/10.1038/nature02900}
}

@article{leemans2006gev,
  title={GeV electron beams from a centimetre-scale accelerator},
  author={Leemans, W. P. and Nagler, B. and Gonsalves, A. J. and T{\'o}th, C. and Nakamura, K. and Geddes, C. G. R. and Esarey, E. and Schroeder, C. B. and Hooker, S. M.},
  journal={Nature physics},
  volume={2},
  number={10},
  pages={696--699},
  year={2006},
  publisher={Nature Publishing Group UK London},
  doi = {10.1038/nphys418},
  url = {https://doi.org/10.1038/nphys418}
}

@article{gorbunov1987excitation,
  title={Excitation of plasma waves by an electromagnetic wave packet},
  author={Gorbunov, LM and Kirsanov, VI},
  journal={Zh. Eksp. Teor. Fiz},
  volume={93},
  pages={509--518},
  year={1987},
  url = {http://www.jetp.ras.ru/cgi-bin/dn/e_066_02_0290.pdf}
}

@article{sprangle1988laser,
  title={Laser wakefield acceleration and relativistic optical guiding},
  author={Sprangle, P and Esarey, E and Ting, A and Joyce, G},
  journal={Applied Physics Letters},
  volume={53},
  number={22},
  pages={2146--2148},
  year={1988},
  publisher={AIP Publishing},
  doi = {10.1063/1.100300},
  url = {https://doi.org/10.1063/1.100300}
}

@article{bulanov1989excitation,
  title={Excitation of ultrarelativistic plasma waves by pulse of electromagnetic radiation},
  author={Bulanov, SV and Kirsanov, VI and Sakharov, AS},
  journal={JETP Lett},
  volume={50},
  number={4},
  pages={176--178},
  year={1989},
  url = {http://jetpletters.ru/ps/1127/article_17078.pdf}
}

@article{berezhiani1990relativistic,
  title={Relativistic wake-field generation by an intense laser pulse in a plasma},
  author={Berezhiani, VI and Murusidze, IG},
  journal={Physics Letters A},
  volume={148},
  number={6-7},
  pages={338--340},
  year={1990},
  publisher={Elsevier},
  doi = {10.1016/0375-9601(90)90813-4},
  url = {https://doi.org/10.1016/0375-9601(90)90813-4} 
}

@article{esarey1993optically,
  title={Optically guided laser wake-field acceleration},
  author={Esarey, Eric and Sprangle, Phillip and Krall, Jonathan and Ting, Antonio and Joyce, Glenn},
  journal={Physics of Fluids B: Plasma Physics},
  volume={5},
  number={7},
  pages={2690--2697},
  year={1993},
  publisher={American Institute of Physics},
  doi = {10.1063/1.860707},
  url = {https://doi.org/10.1063/1.860707}
}

@article{bulanov1995two,
  title={Two-dimensional regimes of self-focusing, wake field generation, and induced focusing of a short intense laser pulse in an underdense plasma},
  author={Bulanov, SV and Pegoraro, Francesco and Pukhov, AM},
  journal={Physical review letters},
  volume={74},
  number={5},
  pages={710},
  year={1995},
  publisher={APS},
  doi = {10.1103/PhysRevLett.74.710},
  url = {https://doi.org/10.1103/PhysRevLett.74.710}
}

@article{bulanov1997transverse,
  title={Transverse-wake wave breaking},
  author={Bulanov, SV and Pegoraro, Francesco and Pukhov, AM and Sakharov, AS},
  journal={Physical review letters},
  volume={78},
  number={22},
  pages={4205},
  year={1997},
  publisher={APS},
  doi = {10.1103/PhysRevLett.78.4205},
  url = {https://doi.org/10.1103/PhysRevLett.78.4205}
}

@article{PhysRevLett.122.084801,
  title = {Petawatt Laser Guiding and Electron Beam Acceleration to 8 GeV in a Laser-Heated Capillary Discharge Waveguide},
  author = {Gonsalves, A. J. and Nakamura, K. and Daniels, J. and Benedetti, C. and Pieronek, C. and de Raadt, T. C. H. and Steinke, S. and Bin, J. H. and Bulanov, S. S. and van Tilborg, J. and Geddes, C. G. R. and Schroeder, C. B. and T\'oth, Cs. and Esarey, E. and Swanson, K. and Fan-Chiang, L. and Bagdasarov, G. and Bobrova, N. and Gasilov, V. and Korn, G. and Sasorov, P. and Leemans, W. P.},
  journal = {Phys. Rev. Lett.},
  volume = {122},
  issue = {8},
  pages = {084801},
  numpages = {6},
  year = {2019},
  month = {Feb},
  publisher = {American Physical Society},
  doi = {10.1103/PhysRevLett.122.084801},
  url = {https://link.aps.org/doi/10.1103/PhysRevLett.122.084801}
}

@article{aniculaesei2024acceleration,
  title={The acceleration of a high-charge electron bunch to 10 GeV in a 10-cm nanoparticle-assisted wakefield accelerator},
  author={Aniculaesei, Constantin and Ha, Thanh and Yoffe, Samuel and Labun, Lance and Milton, Stephen and McCary, Edward and Spinks, Michael M and Quevedo, Hernan J and Labun, Ou Z and Sain, Ritwik and others},
  journal={Matter and Radiation at Extremes},
  volume={9},
  number={1},
  year={2024},
  publisher={AIP Publishing},
  doi = {10.1063/5.0161687},
  url = {https://doi.org/10.1063/5.0161687}
}

@article{picksley2024matched,
  title={Matched guiding and controlled injection in dark-current-free, 10-GeV-class, channel-guided laser-plasma accelerators},
  author={Picksley, A and Stackhouse, J and Benedetti, C and Nakamura, K and Tsai, HE and Li, R and Miao, B and Shrock, JE and Rockafellow, E and Milchberg, HM and others},
  journal={Physical Review Letters},
  volume={133},
  number={25},
  pages={255001},
  year={2024},
  publisher={APS},
  doi = {10.1103/PhysRevLett.133.255001},
  url = {https://doi.org/10.1103/PhysRevLett.133.255001}
}

@article{li2025longitudinal,
  title={Longitudinal tapering in gas jets for increased efficiency of 10-GeV class laser plasma accelerators},
  author={Li, R and Picksley, A and Benedetti, C and Filippi, F and Stackhouse, J and Fan-Chiang, L and Tsai, HE and Nakamura, K and Schroeder, CB and van Tilborg, J and others},
  journal={Review of Scientific Instruments},
  volume={96},
  number={4},
  year={2025},
  publisher={AIP Publishing},
  doi = {10.1063/5.0250698},
  url = {https://doi.org/10.1063/5.0250698}
}

@article{zgadzaj2024plasma,
  title={Plasma electron acceleration driven by a long-wave-infrared laser},
  author={Zgadzaj, R and Welch, J and Cao, Y and Amorim, LD and Cheng, A and Gaikwad, A and Iapozzutto, P and Kumar, P and Litvinenko, VN and Petrushina, I and others},
  journal={Nature Communications},
  volume={15},
  number={1},
  pages={4037},
  year={2024},
  publisher={Nature Publishing Group UK London},
  doi = {10.1038/s41467-024-48413-y},
  url = {https://doi.org/10.1038/s41467-024-48413-y}
}

@article{PhysRevLett.126.104801,
  title = {Bayesian Optimization of a Laser-Plasma Accelerator},
  author = {Jalas, S. and Kirchen, M. and Messner, P. and Winkler, P. and H\"ubner, L. and Dirkwinkel, J. and Schnepp, M. and Lehe, R. and Maier, A. R.},
  journal = {Phys. Rev. Lett.},
  volume = {126},
  issue = {10},
  pages = {104801},
  numpages = {6},
  year = {2021},
  month = {Mar},
  publisher = {American Physical Society},
  doi = {10.1103/PhysRevLett.126.104801},
  url = {https://link.aps.org/doi/10.1103/PhysRevLett.126.104801}
}

@article{PhysRevLett.129.094801,
  title = {Energy Compression and Stabilization of Laser-Plasma Accelerators},
  author = {Ferran Pousa, A. and Agapov, I. and Antipov, S. A. and Assmann, R. W. and Brinkmann, R. and Jalas, S. and Kirchen, M. and Leemans, W. P. and Maier, A. R. and Martinez de la Ossa, A. and Osterhoff, J. and Th\'evenet, M.},
  journal = {Phys. Rev. Lett.},
  volume = {129},
  issue = {9},
  pages = {094801},
  numpages = {8},
  year = {2022},
  month = {Aug},
  publisher = {American Physical Society},
  doi = {10.1103/PhysRevLett.129.094801},
  url = {https://link.aps.org/doi/10.1103/PhysRevLett.129.094801}
}

@article{hue2023control,
  title={Control of electron beam current, charge, and energy spread using density downramp injection in laser wakefield accelerators},
  author={Hue, C{\'e}line S and Wan, Yang and Levine, Eitan Y and Malka, Victor},
  journal={Matter and radiation at extremes},
  volume={8},
  number={2},
  year={2023},
  publisher={AIP Publishing},
  doi = {10.1063/5.0126293},
  url = {https://doi.org/10.1063/5.0126293}
}

@article{habib2023attosecond,
  title={Attosecond-Angstrom free-electron-laser towards the cold beam limit},
  author={Habib, Ahmad F and Manahan, GG and Scherkl, P and Heinemann, T and Sutherland, A and Altuiri, R and Alotaibi, BM and Litos, M and Cary, J and Raubenheimer, T and others},
  journal={Nature communications},
  volume={14},
  number={1},
  pages={1054},
  year={2023},
  publisher={Nature Publishing Group UK London},
  doi = {10.1038/s41467-023-36592-z},
  url = {https://doi.org/10.1038/s41467-023-36592-z}
}

@article{maity2024parametric,
  title={Parametric analysis of electron beam quality in laser wakefield acceleration based on the truncated ionization injection mechanism},
  author={Maity, S. and Mondal, A. and Vishnyakov, E. and Molodozhentsev, A.},
  journal={Plasma Physics and Controlled Fusion},
  volume={66},
  number={3},
  pages={035012},
  year={2024},
  publisher={IOP Publishing},
  doi = {10.1088/1361-6587/ad238e},
  url = {https://doi.org/10.1088/1361-6587/ad238e}
}

@article{guan2025achieving,
  title={Achieving high quality electron beam with ultralow energy spread from mismatched plasma channels},
  author={Guan, Jiabao and Lei, Qiannan and Zhong, Jianhua and Liu, Lanxin and Nie, Yuancun and Xia, Guoxing and Wang, Jike},
  journal={Scientific Reports},
  volume={15},
  number={1},
  pages={11774},
  year={2025},
  publisher={Nature Publishing Group UK London},
  doi = {10.1038/s41598-025-90741-6},
  url = {https://doi.org/10.1038/s41598-025-90741-6}
}

@article{PhysRevLett.110.185006,
  title = {Shock-Front Injector for High-Quality Laser-Plasma Acceleration},
  author = {Buck, A. and Wenz, J. and Xu, J. and Khrennikov, K. and Schmid, K. and Heigoldt, M. and Mikhailova, J. M. and Geissler, M. and Shen, B. and Krausz, F. and Karsch, S. and Veisz, L.},
  journal = {Phys. Rev. Lett.},
  volume = {110},
  issue = {18},
  pages = {185006},
  numpages = {5},
  year = {2013},
  month = {May},
  publisher = {American Physical Society},
  doi = {10.1103/PhysRevLett.110.185006},
  url = {https://link.aps.org/doi/10.1103/PhysRevLett.110.185006}
}

@article{wang2021free,
  title={Free-electron lasing at 27 nanometres based on a laser wakefield accelerator},
  author={Wang, Wentao and Feng, Ke and Ke, Lintong and Yu, Changhai and Xu, Yi and Qi, Rong and Chen, Yu and Qin, Zhiyong and Zhang, Zhijun and Fang, Ming and others},
  journal={Nature},
  volume={595},
  number={7868},
  pages={516--520},
  year={2021},
  publisher={Nature Publishing Group UK London},
  doi = {10.1038/s41586-021-03678-x},
  url = {https://doi.org/10.1038/s41586-021-03678-x}
}

@article{PhysRevLett.126.174801,
  title = {Optimal Beam Loading in a Laser-Plasma Accelerator},
  author = {Kirchen, M. and Jalas, S. and Messner, P. and Winkler, P. and Eichner, T. and H\"ubner, L. and H\"ulsenbusch, T. and Jeppe, L. and Parikh, T. and Schnepp, M. and Maier, A. R.},
  journal = {Phys. Rev. Lett.},
  volume = {126},
  issue = {17},
  pages = {174801},
  numpages = {7},
  year = {2021},
  month = {Apr},
  publisher = {American Physical Society},
  doi = {10.1103/PhysRevLett.126.174801},
  url = {https://link.aps.org/doi/10.1103/PhysRevLett.126.174801}
}

@article{foerster2022stable,
  title={Stable and high-quality electron beams from staged laser and plasma wakefield accelerators},
  author={Foerster, FM and D{\"o}pp, A and Haberstroh, F and Grafenstein, K v and Campbell, D and Chang, Y-Y and Corde, S and Couperus Cabada{\u{g}}, JP and Debus, A and Gilljohann, MF and others},
  journal={Physical Review X},
  volume={12},
  number={4},
  pages={041016},
  year={2022},
  publisher={APS},
  doi = {10.1103/PhysRevX.12.041016},
  url = {https://doi.org/10.1103/PhysRevX.12.041016}
}

@article{winkler2025active,
  title={Active energy compression of a laser-plasma electron beam},
  author={Winkler, P and Trunk, Maximilian and H{\"u}bner, L and Martinez de la Ossa, A and Jalas, S and Kirchen, M and Agapov, I and Antipov, SA and Brinkmann, R and Eichner, T and others},
  journal={Nature},
  pages={1--4},
  year={2025},
  publisher={Nature Publishing Group UK London},
  doi = {10.1038/s41586-025-08772-y},
  url = {https://doi.org/10.1038/s41586-025-08772-y}
}

@inproceedings{walker2017horizon,
  title={Horizon 2020 EuPRAXIA design study},
  author={Walker, P. A. and Alesini, P. D. and Alexandrova, A. S. and Anania, M. P. and Andreev, N. E. and Andriyash, I. and Aschikhin, A. and Assmann, R. W. and Audet, T. and Bacci, A. and others},
  booktitle={Journal of Physics: Conference Series},
  volume={874},
  number={1},
  pages={012029},
  year={2017},
  organization={IOP Publishing}
}

@article{emma2021free,
  title={Free electron lasers driven by plasma accelerators: status and near-term prospects},
  author={Emma, Claudio and Van Tilborg, Jeroen and Assmann, R and Barber, S and Cianchi, A and Corde, S and Couprie, ME and D’arcy, R and Ferrario, M and Habib, AF and others},
  journal={High Power Laser Science and Engineering},
  volume={9},
  pages={e57},
  year={2021},
  publisher={Cambridge University Press},
  doi = {10.1017/hpl.2021.39},
  url = {https://doi.org/10.1017/hpl.2021.39}
}

@inproceedings{vishnyakov2023compact,
  title={Compact undulator-based soft x-ray radiation source at ELI Beamlines: user-oriented program},
  author={Vishnyakov, E. A. and Du Mai, D. and Green, J. T. and Mondal, A. and Jan{\v{c}}{\'a}rek, A. and Zimmermann, P. and Niekrasz, S. and Maity, S. and Molodozhentsev, A. Y.},
  booktitle={Compact Radiation Sources from EUV to Gamma-rays: Development and Applications},
  volume={12582},
  pages={53--62},
  year={2023},
  doi = {10.1117/12.2665377},
  url = {https://doi.org/10.1117/12.2665377},
  organization={SPIE}
}

@inproceedings{molodozhentsev:fls2023-th2c2,
  title = {Development of Laser-Driven Plasma Accelerator Undulator Radiation Source at ELI-Beamlines},
  author       = {Molodozhentsev, A. and Green, J. T. and Mondal, A. and Maity, S. and Niekrasz, S. and Vishnyakov, E. and Jancarek, A. and Zimmermann, P.},
  booktitle    = {Proc. 67th ICFA Adv. Beam Dyn. Workshop Future Light Sources (FLS'23)},
  pages        = {237--240},
  number       = {67},
  publisher    = {JACoW Publishing, Geneva, Switzerland},
  year         = {2024},
  issn         = {2673-7035},
  isbn         = {978-3-95450-224-0},
  doi          = {10.18429/JACoW-FLS2023-TH2C2},
  url          = {http://jacow.org/fls2023/papers/th2c2.pdf}
}

@inproceedings{Whitehead:2024uco,
    title = {Conceptual design of the laser-plasma accelerator based soft X-ray Free Electron Laser},
    author = {Molodozhentsev, A. and Green, J. T. and Rus, B. and Whitehead, A. and Maity, S. and Sasorov, P. and Jancarek, A. and Albrecht, M. and Miceski, M. and Niekrasz, S. and Zimmermann, P.},
    doi = {10.18429/JACoW-IPAC2024-MOPG31},
    journal = {JACoW},
    volume = {IPAC2024},
    pages = {MOPG31},
    year = {2024},
    url  = {https://accelconf.web.cern.ch/ipac2024/pdf/MOPG31.pdf}
}

@inproceedings{Whitehead:2024dkn,
    title = {Plasma accelerator based free electron laser program at ELI-ERIC (ELI-Beamlines)},
    author = {Molodozhentsev, A. and Green, J. T. and Rus, B. and Whitehead, A. and Maity, S. and Sasorov, P. and Jancarek, A. and Albrecht, M. and Miceski, M. and Niekrasz, S. and Zimmermann, P.},
    doi = {10.18429/JACoW-IPAC2024-MOPG32},
    journal = {JACoW},
    volume = {IPAC2024},
    pages = {MOPG32},
    year = {2024},
    url  = {https://accelconf.web.cern.ch/ipac2024/pdf/MOPG32.pdf}
}

@article{zhou2025compact,
  title={Compact dose delivery of laser-accelerated high-energy electron beams toward radiotherapy applications},
  author={Zhou, Bing and Guo, Zhiyuan and Wan, Yang and Liu, Shuang and Peng, Bo and Hua, Jianfei and Lu, Wei},
  journal={Physical Review Accelerators and Beams},
  volume={28},
  number={10},
  pages={101304},
  year={2025},
  publisher={APS},
  doi = {10.1103/xdsm-7xmf},
  url = {https://doi.org/10.1103/xdsm-7xmf}
}

@article{maity2025coupling,
  title={Coupling and acceleration of externally injected electron beams in laser-driven plasma wakefields},
  author={Maity, Srimanta and Sasorov, Pavel and Molodozhentsev, Alexander},
  journal={Journal of Physics D: Applied Physics},
  volume={58},
  number={14},
  pages={145204},
  year={2025},
  publisher={IOP Publishing},
  doi = {10.1088/1361-6463/adb6b9},
  url = {https://doi.org/10.1088/1361-6463/adb6b9}
}

@article{bulanov1992nonlinear,
    author = {Bulanov, S. V. and Inovenkov, I. N. and Kirsanov, V. I. and Naumova, N. M. and Sakharov, A. S.},
    title = "{Nonlinear depletion of ultrashort and relativistically strong laser pulses in an underdense plasma}",
    journal = {Physics of Fluids B: Plasma Physics},
    volume = {4},
    number = {7},
    pages = {1935-1942},
    year = {1992},
    month = {07},
    issn = {0899-8221},
    doi = {10.1063/1.860046},
    url = {https://doi.org/10.1063/1.860046},
    publisher={American Institute of Physics}
}

@article{xu2005electron,
  title={Electron self-injection and acceleration driven by a tightly focused intense laser beam in an underdense plasma},
  author={Xu, H. and Yu, W. and Lu, P. and Senecha, V. K. and He, F. and Shen, B. and Qian, L. and Li, R. and Xu, Z.},
  journal={Physics of plasmas},
  volume={12},
  number={1},
  pages={013105},
  year={2005},
  publisher={American Institute of Physics}
}

@article{PhysRevLett.103.135004,
  title = {Electron Self-Injection and Trapping into an Evolving Plasma Bubble},
  author = {Kalmykov, S. and Yi, S. A. and Khudik, V. and Shvets, G.},
  journal = {Phys. Rev. Lett.},
  volume = {103},
  issue = {13},
  pages = {135004},
  numpages = {4},
  year = {2009},
  month = {Sep},
  publisher = {American Physical Society},
  doi = {10.1103/PhysRevLett.103.135004},
  url = {https://link.aps.org/doi/10.1103/PhysRevLett.103.135004}
}

@article{PhysRevLett.103.215006,
  title = {Measurements of the Critical Power for Self-Injection of Electrons in a Laser Wakefield Accelerator},
  author = {Froula, D. H. and Clayton, C. E. and D\"oppner, T. and Marsh, K. A. and Barty, C. P. J. and Divol, L. and Fonseca, R. A. and Glenzer, S. H. and Joshi, C. and Lu, W. and Martins, S. F. and Michel, P. and Mori, W. B. and Palastro, J. P. and Pollock, B. B. and Pak, A. and Ralph, J. E. and Ross, J. S. and Siders, C. W. and Silva, L. O. and Wang, T.},
  journal = {Phys. Rev. Lett.},
  volume = {103},
  issue = {21},
  pages = {215006},
  numpages = {4},
  year = {2009},
  month = {Nov},
  publisher = {American Physical Society},
  doi = {10.1103/PhysRevLett.103.215006},
  url = {https://link.aps.org/doi/10.1103/PhysRevLett.103.215006}
}

@article{PhysRevSTAB.15.011302,
  title = {Self-injection threshold in self-guided laser wakefield accelerators},
  author = {Mangles, S. P. D. and Genoud, G. and Bloom, M. S. and Burza, M. and Najmudin, Z. and Persson, A. and Svensson, K. and Thomas, A. G. R. and Wahlstr\"om, C.-G.},
  journal = {Phys. Rev. ST Accel. Beams},
  volume = {15},
  issue = {1},
  pages = {011302},
  numpages = {6},
  year = {2012},
  month = {Jan},
  publisher = {American Physical Society},
  doi = {10.1103/PhysRevSTAB.15.011302},
  url = {https://link.aps.org/doi/10.1103/PhysRevSTAB.15.011302}
}

@article{PhysRevE.58.R5257,
  title = {Particle injection into the wave acceleration phase due to nonlinear wake wave breaking},
  author = {Bulanov, S. and Naumova, N. and Pegoraro, F. and Sakai, J.},
  journal = {Phys. Rev. E},
  volume = {58},
  issue = {5},
  pages = {R5257--R5260},
  numpages = {0},
  year = {1998},
  month = {Nov},
  publisher = {American Physical Society},
  doi = {10.1103/PhysRevE.58.R5257},
  url = {https://link.aps.org/doi/10.1103/PhysRevE.58.R5257}
}

@article{PhysRevLett.100.215004,
  title = {Plasma-Density-Gradient Injection of Low Absolute-Momentum-Spread Electron Bunches},
  author = {Geddes, C. G. R. and Nakamura, K. and Plateau, G. R. and Toth, C. and Cormier-Michel, E. and Esarey, E. and Schroeder, C. B. and Cary, J. R. and Leemans, W. P.},
  journal = {Phys. Rev. Lett.},
  volume = {100},
  issue = {21},
  pages = {215004},
  numpages = {4},
  year = {2008},
  month = {May},
  publisher = {American Physical Society},
  doi = {10.1103/PhysRevLett.100.215004},
  url = {https://link.aps.org/doi/10.1103/PhysRevLett.100.215004}
}

@article{ke2021near,
  title={Near-GeV electron beams at a few per-mille level from a laser wakefield accelerator via density-tailored plasma},
  author={Ke, LT and Feng, K and Wang, WT and Qin, ZY and Yu, CH and Wu, Y and Chen, Y and Qi, R and Zhang, ZJ and Xu, Y and others},
  journal={Physical review letters},
  volume={126},
  number={21},
  pages={214801},
  year={2021},
  publisher={APS},
  doi = {10.1103/PhysRevLett.126.214801},
  url = {https://doi.org/10.1103/PhysRevLett.126.214801}
}

@article{PhysRevLett.76.2073,
  title = {Laser Injection of Ultrashort Electron Pulses into Wakefield Plasma Waves},
  author = {Umstadter, D. and Kim, J. K. and Dodd, E.},
  journal = {Phys. Rev. Lett.},
  volume = {76},
  issue = {12},
  pages = {2073--2076},
  numpages = {0},
  year = {1996},
  month = {Mar},
  publisher = {American Physical Society},
  doi = {10.1103/PhysRevLett.76.2073},
  url = {https://link.aps.org/doi/10.1103/PhysRevLett.76.2073}
}

@article{PhysRevLett.79.2682,
  title = {Electron Injection into Plasma Wakefields by Colliding Laser Pulses},
  author = {Esarey, E. and Hubbard, R. F. and Leemans, W. P. and Ting, A. and Sprangle, P.},
  journal = {Phys. Rev. Lett.},
  volume = {79},
  issue = {14},
  pages = {2682--2685},
  numpages = {0},
  year = {1997},
  month = {Oct},
  publisher = {American Physical Society},
  doi = {10.1103/PhysRevLett.79.2682},
  url = {https://link.aps.org/doi/10.1103/PhysRevLett.79.2682}
}

@article{faure2006controlled,
  title={Controlled injection and acceleration of electrons in plasma wakefields by colliding laser pulses},
  author={Faure, J. and Rechatin, C. and Norlin, A. and Lifschitz, A. and Glinec, Y. and Malka, V.},
  journal={Nature},
  volume={444},
  number={7120},
  pages={737--739},
  year={2006},
  publisher={Nature Publishing Group UK London},
  doi = {10.1038/nature05393},
  url = {https://doi.org/10.1038/nature05393}
}

@article{wang2022injection,
  title={Injection induced by coaxial laser interference in laser wakefield accelerators},
  author={Wang, Jia and Zeng, Ming and Li, Dazhang and Wang, Xiaoning and Lu, Wei and Gao, Jie},
  journal={Matter and Radiation at Extremes},
  volume={7},
  number={5},
  year={2022},
  publisher={AIP Publishing},
  doi = {10.1063/5.0101098},
  url = {https://doi.org/10.1063/5.0101098}
}

@article{bohlen2023colliding,
  title={Colliding pulse injection of polarized electron bunches in a laser-plasma accelerator},
  author={Bohlen, Simon and Gong, Zheng and Quin, Michael J and Tamburini, Matteo and P{\~o}der, Kristjan},
  journal={Physical Review Research},
  volume={5},
  number={3},
  pages={033205},
  year={2023},
  publisher={APS},
  doi = {10.1103/PhysRevResearch.5.033205},
  url = {https://doi.org/10.1103/PhysRevResearch.5.033205}
}

@article{PhysRevLett.104.025003,
  title = {Injection and Trapping of Tunnel-Ionized Electrons into Laser-Produced Wakes},
  author = {Pak, A. and Marsh, K. A. and Martins, S. F. and Lu, W. and Mori, W. B. and Joshi, C.},
  journal = {Phys. Rev. Lett.},
  volume = {104},
  issue = {2},
  pages = {025003},
  numpages = {4},
  year = {2010},
  month = {Jan},
  publisher = {American Physical Society},
  doi = {10.1103/PhysRevLett.104.025003},
  url = {https://link.aps.org/doi/10.1103/PhysRevLett.104.025003}
}

@article{PhysRevLett.104.025004,
  title = {Ionization Induced Trapping in a Laser Wakefield Accelerator},
  author = {McGuffey, C. and Thomas, A. G. R. and Schumaker, W. and Matsuoka, T. and Chvykov, V. and Dollar, F. J. and Kalintchenko, G. and Yanovsky, V. and Maksimchuk, A. and Krushelnick, K. and Bychenkov, V. Yu. and Glazyrin, I. V. and Karpeev, A. V.},
  journal = {Phys. Rev. Lett.},
  volume = {104},
  issue = {2},
  pages = {025004},
  numpages = {4},
  year = {2010},
  month = {Jan},
  publisher = {American Physical Society},
  doi = {10.1103/PhysRevLett.104.025004},
  url = {https://link.aps.org/doi/10.1103/PhysRevLett.104.025004}
}

@article{PhysRevLett.105.105003,
  title = {Self-Guided Laser Wakefield Acceleration beyond 1 GeV Using Ionization-Induced Injection},
  author = {Clayton, C. E. and Ralph, J. E. and Albert, F. and Fonseca, R. A. and Glenzer, S. H. and Joshi, C. and Lu, W. and Marsh, K. A. and Martins, S. F. and Mori, W. B. and Pak, A. and Tsung, F. S. and Pollock, B. B. and Ross, J. S. and Silva, L. O. and Froula, D. H.},
  journal = {Phys. Rev. Lett.},
  volume = {105},
  issue = {10},
  pages = {105003},
  numpages = {4},
  year = {2010},
  month = {Sep},
  publisher = {American Physical Society},
  doi = {10.1103/PhysRevLett.105.105003},
  url = {https://link.aps.org/doi/10.1103/PhysRevLett.105.105003}
}

@article{steyn2025observation,
  title={Observation of laser plasma accelerated electrons with transverse momentum spread below the thermal level},
  author={Steyn, TL and Panchal, A and Vasilovici, O and Sch{\"o}bel, S and Ufer, P and Herrmann, FM and Chang, Y-Y and Moulanier, I and Masckala, M and Khomyshyn, O and others},
  journal={arXiv preprint arXiv:2506.18047},
  year={2025},
  url = {https://arxiv.org/abs/2506.18047}
}

@article{irman2018improved,
  title={Improved performance of laser wakefield acceleration by tailored self-truncated ionization injection},
  author={Irman, A. and Couperus, J. P. and Debus, A. and K{\"o}hler, A. and Kr{\"a}mer, J. M. and Pausch, R. and Zarini, O. and Schramm, U.},
  journal={Plasma Physics and Controlled Fusion},
  volume={60},
  number={4},
  pages={044015},
  year={2018},
  publisher={IOP Publishing},
  doi = {10.1088/1361-6587/aaaef1},
  url = {https://doi.org/10.1088/1361-6587/aaaef1}
}

@book
{
  greenshields2025,
  title     = "OpenFOAM v13 User Guide",
  author    = "Greenshields, Christopher",
  year      = 2025,
  url       = "https://doc.cfd.direct/openfoam/user-guide-v13",
  publisher = "The OpenFOAM Foundation",
  address   = "London, UK"
}

@book{chapman1990mathematical,
  title={The mathematical theory of non-uniform gases: an account of the kinetic theory of viscosity, thermal conduction and diffusion in gases},
  author={Chapman, Sydney and Cowling, Thomas George},
  year={1990},
  publisher={Cambridge university press}
}

@article{bagdasarov2022discharge,
  title={Discharge plasma formation in square capillary with gas supply channels},
  author={Bagdasarov, GA and Kruchinin, KO and Molodozhentsev, A Yu and Sasorov, PV and Bulanov, SV and Gasilov, VA},
  journal={Physical Review Research},
  volume={4},
  number={1},
  pages={013063},
  year={2022},
  publisher={APS},
  doi = {10.1103/PhysRevResearch.4.013063},
  url = {https://doi.org/10.1103/PhysRevResearch.4.013063}
}

@article{derouillat2018smilei,
  title={Smilei: A collaborative, open-source, multi-purpose particle-in-cell code for plasma simulation},
  author = {J. Derouillat and A. Beck and F. Pérez and T. Vinci and M. Chiaramello and A. Grassi and M. Flé and G. Bouchard and I. Plotnikov and N. Aunai and J. Dargent and C. Riconda and M. Grech},
  journal={Computer Physics Communications},
  volume={222},
  pages={351--373},
  year={2018},
  publisher={Elsevier},
  doi = {10.1016/j.cpc.2017.09.024},
  url = {https://doi.org/10.1016/j.cpc.2017.09.024}
}

@article{ammosov1986tunnel,
  title={Tunnel ionization of complex atoms and of atomic ions in an alternating electromagnetic field},
  author={Ammosov, Maxim V and Delone, Nikolai B and Krainov, Vladimir P},
  journal={Soviet Journal of Experimental and Theoretical Physics},
  volume={64},
  number={6},
  pages={1191},
  year={1986},
  url = {http://jetp.ras.ru/cgi-bin/dn/e_064_06_1191},
  publisher = {Maik Nauka/Interperiodica (Russia)}
}

@article{lu2007generating,
  title={Generating multi-GeV electron bunches using single stage laser wakefield acceleration in a 3D nonlinear regime},
  author={Lu, Wei and Tzoufras, M and Joshi, C and Tsung, FS and Mori, WB and Vieira, J and Fonseca, RA and Silva, LO},
  journal={Physical Review Special Topics—Accelerators and Beams},
  volume={10},
  number={6},
  pages={061301},
  year={2007},
  publisher={APS},
  doi = {10.1103/PhysRevSTAB.10.061301},
  url = {https://doi.org/10.1103/PhysRevSTAB.10.061301}
}

@article{10.1063/5.0286730,
    author = {Biagioni, A. and Bourgeois, N. and Brandi, F. and Cassou, K. and Corner, L. and Crincoli, L. and Cros, B. and Dobosz Dufrénoy, S. and Douillet, D. and Drobniak, P. and Faure, J. and Gatti, G. and Grittani, G. and Lorenz, S. and Jones, H. and Lucas, B. and Massimo, F. and Mercier, B. and Molodozhentsev, A. and Monzac, J. and Pattathil, R. and Sarri, G. and Sasorov, P. and Shalloo, R. J. and Steyn, L. and Streeter, M. J. V. and Symes, D. and Thaury, C. and Vernier, A. and Wood, J. C.},
    title = {Technical status report on plasma components and systems in the context of EuPRAXIA},
    journal = {Physics of Plasmas},
    volume = {32},
    number = {11},
    pages = {110501},
    year = {2025},
    month = {11},
    doi = {10.1063/5.0286730},
    url = {https://doi.org/10.1063/5.0286730},
    publisher = {AIP Publishing}
}

@article{kruchinin2025experimental,
  title={Experimental characterization of discharge plasma dynamics in a square capillary for prospective applications in laser wakefield acceleration},
  author={Kruchinin, KO and Mondal, A and Sasorov, PV and Zimmermann, P and Niekrasz, S and Molodozhentsev, A Yu},
  journal={Physics of Plasmas},
  volume={32},
  number={4},
  year={2025},
  publisher={AIP Publishing},
  doi = {10.1063/5.0260100},
  url = {https://doi.org/10.1063/5.0260100}
}

@article{thaury2015shock,
  title={Shock assisted ionization injection in laser-plasma accelerators},
  author={Thaury, C{\'e}dric and Guillaume, E and Lifschitz, Agustin and Ta Phuoc, K and Hansson, Martin and Grittani, G and Gautier, J and Goddet, J-P and Tafzi, A and Lundh, Olle and others},
  journal={Scientific reports},
  volume={5},
  number={1},
  pages={16310},
  year={2015},
  publisher={Nature Publishing Group UK London},
  doi = {10.1038/srep16310},
  url = {https://doi.org/10.1038/srep16310}
}

\end{document}